\documentclass[aps,preprint,superscriptaddress,amsmath,amssymb]{revtex4}
\usepackage{graphicx}
\usepackage{color}
\usepackage{slashed}
\usepackage{subfigure,bbold}

\def\be{\begin{eqnarray}}
\def\ed{\end{eqnarray}}
\def\non{\nonumber}

\begin{document}

{\begin{flushright}{KIAS-P16026}
\end{flushright}}

\title{ Single production of  $X_{\pm 5/3}$ and $Y_{\mp 4/3}$ vector-like quarks  at the LHC}

\author{ Chuan-Hung Chen \footnote{Email: physchen@mail.ncku.edu.tw} }
\affiliation{Department of Physics, National Cheng Kung University, Tainan, Taiwan }

\author{ Takaaki Nomura \footnote{Email: nomura@kias.re.kr }}
\affiliation{School of Physics, Korea Institute for Advanced Study, Seoul 130-722, Republic of Korea}

\date{\today}

\begin{abstract}

Two triplet vector-like quarks (VLQs) with hypercharges of $Y=2/3, -1/3$ and one singlet scalar boson are embedded in the standard model (SM) to resolve the 750 GeV diphoton excess.  The constraints  on the tree-level Higgs- and $Z$-mediated flavor-changing neutral currents  are discussed  in detail.  Besides the resolution of excess,  it is found that  the signal strength of diphoton Higgs decay can have a $10\%$ deviation from the SM prediction and that the upper limits of the branching ratios for rare top-quark decays are $BR(t\to c (h, Z)) < (6.8, 0.48) \times 10^{-5}$. We find that  the production cross section of a single VLQ by electroweak processes is larger than that of VLQ-pair by QCD processes. To explore the  signals of the heavy VLQs at the LHC,  we throughly analyze the production of  single $X_{\pm 5/3}$ and $Y_{\mp 4/3}$ via  $q_i q'_j$ annihilations in $pp$ collisions at $\sqrt{s}=13$ TeV.  It is found that the electroweak production cross sections  for  $d X_{5/3}$, $u Y_{-4/3} $, and $d Y_{4/3} $ channels with $m_X=m_Y=1$ TeV can be  $84.3$, $72.3$, and $157.8$ fb, respectively; and the  dominant decay modes are $X_{5/3} \to ( c, t) W^+$ and $Y_{-4/3} \to (s, b) W^-$. With adopting kinematic cuts, the significance for $pp\to d W^+ t$ channel can be over $5\sigma$.

\end{abstract}

\maketitle

\section{Introduction} 

With the discovery  of  the standard model (SM) Higgs in the ATLAS~\cite{:2012gk} and CMS~\cite{:2012gu} experiments, we have taken one step further toward understanding the electroweak symmetry breaking (EWSB) through the spontaneous symmetry breaking (SSB) mechanism in the scalar sector. The  next mission for the High-Luminosity Large Hadron Collider (LHC) is to explore not only the detailed properties of the SM Higgs, but also the new physics effects.   

Since  problems related to  the origin of neutrino mass, dark matter (DM), and matter-antimatter asymmetry cannot be resolved in the SM, it is believed that the SM of particle physics is an effective theory at the electroweak scale. If new physics exists at the TeV scale, the LHC can detect it. Some potential events indicating the existence of new effects indeed have been observed  in the recent ATLAS and CMS experiments. For instance,  diboson excess of $VV$ with $V=W/Z$ at around 2 TeV was shown by ATLAS~\cite{Aad:2015owa} and CMS~\cite{Khachatryan:2014hpa}; the branching ratio (BR) for lepton-flavor-violating Higgs decay $h\to \mu \tau$ with a $2.4\sigma$ significance  was presented by CMS~\cite{Khachatryan:2015kon}; a resonance at a mass of  750 GeV in the diphoton invariant mass spectrum was reported by  ATLAS~\cite{ATLAS-CONF-2015-081,Aaboud:2016tru} and CMS~\cite{CMS:2015dxe,Khachatryan:2016hje}. Although the results are not conclusive yet, these experimental  measurements have inspired theorists to speculate various effects to interpret the excesses. 
 
Ever since the SM Higgs was observed, the Higgs measurements have approached to the precision level. It becomes an important issue to uncover the physics beyond the SM through the Higgs portal. Precise Higgs measurements can also give strict bounds on the new couplings; for instance,  $\cos(\beta-\alpha)$ in the two-Higgs-doublet model has been limited to be close to the decoupling limit~\cite{Arhrib:2015maa}, and the SM with a fourth generation of chiral fermions has become disfavored~\cite{Eberhardt:2012sb}. 

Although the extension of the SM with chiral fermions has been severely limited, the constraint on the vector-like quark (VLQ) models may not have the same situation due to the use of different representations and coupling structures. Unlike chiral fermion models, where the appearance of chiral quarks has to  accompany chiral leptons due to gauge anomaly,   the gauge anomaly in VLQ models is cancelled automatically. Therefore, it is not necessary to introduce the exotic heavy leptons into the SM when VLQs are added.  Due to their interesting properties, the phenomena of some specific VLQs at the LHC have been investigated from a theoretical viewpoint~\cite{delAguila:2000rc,Atre:2008iu,Mrazek:2009yu,Cacciapaglia:2010vn,Gopalakrishna:2011ef,Botella:2012ju,Okada:2012gy,Cai:2012ji,Cacciapaglia:2012dd,Atre:2013ap,Aguilar-Saavedra:2013qpa,Gopalakrishna:2013hua,Alok:2014yua,Karabacak:2014nca,Cacciapaglia:2015ixa,Beauceron:2014ila,Alok:2015iha,Vignaroli:2015ama,Chen:2015cfa,Angelescu:2015kga,Benbrik:2015fyz}.  In experiments, single VLQs and pairs of VLQs have been produced  at ATLAS~\cite{ATLAS:2012qe,Aad:2014efa,Aad:2015kqa,Aad:2015mba,Aad:2015gdg,Aad:2015tba,Aad:2015voa,Aad:2016shx} and CMS~\cite{Chatrchyan:2013uxa,Chatrchyan:2013wfa,Khachatryan:2015gza,Khachatryan:2015oba,Khachatryan:2015axa}.

Based on the SM gauge symmetry  $SU(2)_L\times U(1)_Y$, the  representations of VLQs  can basically be any $SU(2)$ multiplets.  However, in order to consider the VLQ decays, the possible representations of VLQ couplings to the SM quarks are singlet, doublet, and triplet. To interpret  the excesses  of dibosons and  diphotons indicated  by ATLAS and CMS, we  proposed a model that contains one Higgs singlet and  two triplet VLQs with hypercharges of $Y=2/3$ and $Y=-1/3$, respectively~\cite{Chen:2015cfa,Benbrik:2015fyz}. Since the representations of the VLQs are different from those of the SM quarks, the Higgs- and $Z$-mediated flavor-changing neutral currents (FCNCs) are induced at the tree level and the Cabibbo-Kobayashi-Maskawa (CKM) matrix is non-unitary matrix. In our earlier studies, besides the resolutions of the excesses,  we focused on the leading effects, which were from the left-handed flavor-mixing matrices, and  found that they led to   interesting  contributions to  top FCNCs $t\to q (h, Z)$ and the SM Higgs production and decays.

 In this study, we systematically discuss the left- and right-handed flavor-mixing effects together. We revisit the  constraints and present the bounds from $\Delta F=2$ processes in detail.  With the values of constrained parameters, it is found that the modified top coupling  to the SM Higgs, which arises  from the right-handed flavor mixing, can diminish  the influence of the SM Higgs production and  the decay to diphotons  by around $10\%$ and $-2\%$ deviations from the SM results, respectively. We demonstrate how the changes of the SM CKM matrix elements can be smeared so that the severe bounds from the current measurements of the CKM matrix elements are satisfied~\cite{PDG2014}.  

In addition to the phenomena in flavor physics, we also investigate the single and pair production of VLQs in this work. In the proposed model, the new quarks are $T_{1,2}$, $B_{1,2}$, $X$, and $Y$, whose associated electric charges  are $ 2/3$, $-1/3$, $ 5/3$, and $- 4/3$, respectively. Therefore, $T_{1,2}$ and $B_{1,2}$ can be regarded as top and bottom partners, respectively. Since VLQs $X$ and $Y$ carry the unusual electric charges, they do not have  FCNC couplings to the SM quarks. As a result, their single production and decays are only through charged weak interactions. Since the top and bottom partners involve more complicated FCNC interactions, we concentrate the study on VLQs $X$ and $Y$. It is found that the single production cross sections of $X$ and $Y$ can be larger than the pair production cross sections, which are dominant from quantum chromodynamics (QCD). In order to understand this phenomenon, we analyze  each process  $q_i q'_j \to (X, Y) q_k$ for the single production of $X$ and $Y$.   It is found that with $m_{X(Y)}\sim 1$ TeV, the cross sections for $X d$ and $\bar Y d$ modes can be of  the order of 100 fb while the pair production cross sections are smaller by a factor of around 2.  We postpone the study of event simulation to another paper. 

The rest of this paper is organized as follows. We establish the model, discuss the new flavor mixing effects, and derive the new Higgs and gauge couplings in Section II.  We present the constraints from low-energy and Higgs measurements in Section III. We also study the implications on  top-quark FCNC processes $t\to q (h, Z)$. In Section IV, we discuss the single and pair production for VLQs $X_{5/3}$ and $Y_{-4/3}$, and thoroughly analyze the production mechanism in $pp$ collisions. The conclusions are given in Section V. 

\section{Model and flavor mixings }

\subsection{Model and new interactions}

We extend the SM by including one real Higgs singlet and two vector-like triplet quarks (VLTQs), where the representations of VLTQs in $SU(3)_c\times SU(2)_L \times U(1)_Y$ gauge symmetry are chosen as $(3,3)_{2/3}$ and $(3,3)_{-1/3}$ \cite{Chen:2015cfa}.  For suppressing the mixing between Higgs singlet and doublet,  we impose a $Z_2$ discrete symmetry on the scalar potential, where  the scalar fields follow the transformations $S \to -S$ and $H\to H$ under the $Z_2$ transformation.  Thus, the  scalar potential is expressed as:
 \begin{align}
 V(H, S) &= \mu^2 H^\dagger H + \lambda_1 (H^\dagger H)^2 + m^2_S S^2   + 
  \lambda_2  S^4 + \lambda_3 S^2 (H^\dagger H)\,. \label{eq:VHS}
 \end{align}
We adopt the following representation of $H$:
  \be
H= \left(\begin{array}{cc}
 G^+    \\
\frac{1}{\sqrt{2}} ( v+ h + iG^0)     
\end{array}
  \right)\,,  
  \ed
where $G^+$ and $G^0$ are Goldstone bosons, $h$ is the SM Higgs field, and $v $ is the vacuum expectation value (VEV) of $H$.  The $S$ field cannot develop a non-vanishing VEV in the scalar potential of Eq.~(\ref{eq:VHS})  when $\lambda_{2,3}>0$. Due to the $Z_2$ symmetry,  $h$ and $S$ do not mix at the tree level; thus  $m_S$ is the mass of $S$, $v=\sqrt{-\mu^2/\lambda_1}$, and $m_h= \sqrt{2 \lambda_1} v \approx 125$ GeV is the mass of the SM Higgs~\cite{:2012gk,:2012gu}.   We note that the $Z_2$ symmetry is  softly broken by some other sector of Lagrangian.

 The gauge-invariant Yukawa couplings of VLTQs to the SM quarks,  the SM Higgs doublet, and  the new Higgs singlet are expressed as: 
\be
-{\cal L}^{Y}_{\rm VLTQ} &=&  \bar Q_L {\bf Y_1} F_{1R} \tilde{H}  + \bar Q_L {\bf Y_2} F_{2R} H+   y_{1}  Tr(\bar F_{1L}  F_{1R} ) S
+  y_2  Tr(\bar F_{2L} F_{2R} ) S \non \\
&+& M_{F_1} Tr(\bar F_{1L}  F_{1R} ) + M_{F_2} Tr(\bar F_{2L}  F_{2R})+ H.c.\,,  \label{eq:yukawa}
\ed
where $Q_L$ is the left-handed SM quark doublet, all flavor indices are hidden,  $\tilde H =i \tau_2 H^*$, and   $F_{1(2)}$ is the $2\times 2$ VLTQ with hypercharge $2/3(-1/3)$,  whose  representations of  $F_{1,2}$ are:
  \be
F_{1} = 
\left(
\begin{array}{cc}
 U_1/\sqrt{2} & X    \\
 D_1 &  -U_1/\sqrt{2}     
\end{array}
\right)\,, \  F_{2} = 
\left(
\begin{array}{cc}
 D_2/\sqrt{2} & U_2    \\
 Y &  -D_2/\sqrt{2}     
\end{array}
\right)\,.
\ed
Under $Z_2$ transformation, $F_{1L, 2L}\to - F_{1L,2L}$. The electric charges of $U_{1,2}$, $D_{1,2}$, $X$, and $Y$ are $2/3$, $-1/3$, $5/3$, and $-4/3$, respectively. Therefore, $U_{1,2} (D_{1,2})$ can mix with up (down)-type SM quarks.  The masses of VLTQs do not originate  from the electroweak symmetry breaking. Due to the gauge symmetry, VLTQs in a given multiplet state are degenerate, and denoted by $M_{F_{1(2)}}$.  Since the mass terms of VLTQs do not involve the $S$ field and the associated operators are dimension-3, the discrete $Z_2$ symmetry is softly broken by $M_{F_{1,2}}$ terms. 

It is worth mentioning  that ${\bf Y}_{1,2}$ results in the mixture of the SM quarks and VLTQs; consequently, the $h$ coupling to the top quark is modified and the $h$ couplings to VLTQs are induced. As a result, the SM Higgs production  cross section via gluon-gluon fusion (ggF) and its decays will be modified.  In the next subsection, we discuss the modifications in detail. 
We note that the $Z_2$ breaking effects will induce $SH^\dagger H$ term through one-loop diagrams in the scalar potential.  However, in addition to the suppression  factor $1/(4\pi)^2$, the loop effects are suppressed by the small Yukawa couplings ${\bf Y}_{1,2}$ ( see the detailed  discussions later). As a result, the induced VEV of $S$ field and the induced mixing between $h$ and $S$ are small,  and the BR for $S\to hh$ decay is a factor of two smaller than   that for $S\to \gamma\gamma$ decay ~\cite{Benbrik:2015fyz}.

Next, we discuss the weak interactions of VLTQs.  As usual, we write the covariant derivative of $SU(2)_L\times U(1)_Y$ as:
  \be
 D_\mu = \partial_\mu + i \frac{g}{\sqrt{2}} \left( T^+ W^+_\mu +T^- W^{-}_\mu \right) + i\frac{g}{c_W} \left(   T^3 -  s^2_W Q \right) Z_\mu+ i e Q A_\mu \,,
 \ed
where $W^{\pm}_\mu$, $Z_\mu$, and $A_\mu$ are the gauge bosons in the SM, $g$ is the gauge coupling of $SU(2)_L$, $s_W(c_W)=\sin\theta_W (\cos\theta_W)$, $\theta_W$ is the Weinberg angle, $T^{\pm} =T^1 \pm i T^2$, and the charge operator $Q = T^3 + Y$, where   $Y$ is the hypercharge of the particle. Thus, the gauge interactions of VLTQs are written as~\cite{Chen:2015cfa}:
 \be
 {\cal L}_{VFF} &=& -g \left[\left( \bar X \gamma^\mu U_1 + \bar U_1 \gamma^\mu D_1 + \bar D_2 \gamma^\mu Y + \bar U_2 \gamma^\mu D_2 \right) W^+_\mu + h.c. \right] \non \\
 &-& \left[ \frac{g}{c_W}  \bar F_1 \gamma^\mu \left( T^3 -s^2_W Q_1 \right) F_1 Z_\mu + e \bar F_1 \gamma^\mu Q_1 F_1 A_\mu  + ( F_1 \to F_2, Q_1 \to Q_2 ) \right]\,, \label{eq:VFF}
 \ed
where we express the triplet VLQs as  $F^T_1 = (X, U_1, D_1)$ and $F^T_2 = (U_2, D_2, Y)$, diag$T^3=(1,0,-1)$, diag$Q_1= (5/3, 2/3, -1/3)$, and diag$Q_2 = (2/3, -1/3, -4/3)$.  To  further understand the weak interactions in terms of physical states, we have to investigate the structures of flavor mixings when ${\bf Y}_{1,2}$ effects in Eq.~(\ref{eq:yukawa}) are involved.

\subsection{Flavor mixings and Higgs-mediated FCNCs }

The introduced two $SU(2)_L$ triplet VLQs contain the quarks with electric charges of $+2/3$ and $-1/3$. From the new  Higgs Yukawa couplings to  the SM Higgs and VLTQs,   the mixture between the SM quarks and VLTQs is generated after EWSB. In order to get   the physical mass eigenstates of quarks and the new flavor mixings, we have to diagonalize the mass matrices of the SM quarks and VLTQs. With the Yukawa couplings $Y_{1i}$ and $Y_{2i}$ (i=1-3) in Eq.~(\ref{eq:yukawa}), the quark mass terms  are given by:
  \be
  -{\cal L}_{\rm mass} = \bar {\bf q}_L {\bf m}_q {\bf q}_R + \bar {\bf q}_L {\bf Y}^q v {\bf F}_R  + \bar {\bf F}_L {\bf m}_F {\bf F}_R + H.c.\,,  
  \ed
 where ${\bf q}^T=(u,c,t)$ or $(d,s,b)$ denotes the SM up- or down-type quarks. We have chosen the basis such that ${\bf m}_q$ is a $3\times 3$ diagonal matrix, ${\bf F}^T = (U_1, U_2)$ or $(D_1, D_2)$ is the VLTQ with charge $+2/3$ or $-1/3$ , diag${\bf m}_F = (m_{F_1}, m_{F_2}$), and 
 \be
 {\bf Y^u}=\left(\begin{array}{cc} 
                                  Y_{11}/2 & Y_{21}/\sqrt{2} \\
                                   Y_{12}/2 & Y_{22}/\sqrt{2} \\
                                    Y_{13}/2 & Y_{23}/\sqrt{2} \end{array} \right)\,, ~
   {\bf Y^d}=\left(\begin{array}{cc} 
                                  Y_{11}/\sqrt{2} & -Y_{21}/2 \\
                                   Y_{12}/\sqrt{2} & -Y_{22}/2\\
                                    Y_{13}/\sqrt{2} & -Y_{23}/2\end{array} \right)\,. \label{eq:Yud}
 \ed
 We do not have $\bar F_L q_R$ terms due to  gauge invariance. 
 The quark mass matrices for electric charges $+2/3$ and $-1/3$ now become $5 \times 5$ matrices. One can introduce the $5\times 5$ unitary matrices $V^q_L$ and $V^q_R$ to diagonalize the mass matrices, namely $M^{\rm dia}_{q} = V^q_L M_{q} V^{q\dagger}_{R}$. In order to obtain the information of $V^q_{L, R}$, we consider the multiplications of mass matrices to be $M^{\rm dia}_q M^{\rm dia \dagger} = V^q_L M_q M^\dagger_q V^{q \dagger}_L$ and $M^{\rm dia \dagger}_q M^{\rm dia}_q = V^q_R M^\dagger_q M_q V^{q\dagger}_R$, where $M_q M^\dagger_q$ and $M^\dagger_q M_q$ are expressed as:
\begin{align}
M_q M^\dagger_q &= \left(
\begin{array}{cc}
 {\bf m}_q {\bf m}^\dagger_q + v^2 {\bf Y}^q {\bf Y}^{q\dagger} &  v {\bf Y}^q {\bf m}^\dagger_F  \\ 
{\bf m}_F {\bf Y}^{q\dagger} v &   {\bf m}_F {\bf m}^\dagger_F 
\end{array}
\right)\,,  \non\\
M^\dagger_q M_q& = \left(
\begin{array}{cc}
 {\bf m}^\dagger_q {\bf m}_q  &  v {\bf m}^\dagger_q {\bf Y}^q   \\ 
 v{\bf Y}^{q\dagger}  {\bf m}_q &   {\bf m}^\dagger_F {\bf m}_F + v^2 {\bf Y}^{q\dagger} {\bf Y}^{q}
\end{array}
\right) \label{eq:mass2}  
 \end{align} 
with ${\bf m}^\dagger_q ={\bf m}_q$, and ${\bf m}^\dagger_{F} = \bf{m}_{F}$.  It is clear that  the off-diagonal matrix elements in $M_q M^\dagger_q$ are related to $v {\bf Y}^q {\bf m}_F$ while those in $M^\dagger_q M_q$ are $v {\bf m}_q {\bf Y}^q$. Due to $m_{q_j}, v Y^q_{ij} \ll m_{F_i}$,  the unitary matrices $V^q_{L,R}$ can be expanded with respect to $v /m_{F_i}$ and $m_{q}/Mm_{F_i}$; at the leading order approximation, they can be formulated as:
  \be
  V^q_\chi \approx  \left( \begin{array}{cc} 
                                     \mathbb{1}_{3\times 3} & -(\pmb{ \varepsilon}^q_\chi)_{3\times 2} \\
                                     (\pmb{ \varepsilon}^{q \dagger}_\chi)_{2\times 3} &  \mathbb{1}_{2\times 2}
                                     \end{array} \right)\,, \label{eq:VqLR}
  \ed
  where $\chi=L, R$, $\pmb{ \varepsilon}^q_R = v {\bf m}_q {\bf Y}^q/{\bf m}^2_F$, and $\pmb{ \varepsilon}^q_L = v {\bf Y}^q {\bf m}_F /{\bf m}^2_F$. We find that the effects of $\pmb{\varepsilon}^q_R$ are suppressed by $v m_q /m^2_{F_i}$ while those of $\pmb{\varepsilon}^q_L$ are associated with $v/m_{F_i}$. Since the top and bottom quarks are much heavier than other SM quarks, in this study we keep the contributions from  $\varepsilon^{u(d)}_{R13}= v m_{t(b)} Y^{u(d)}_{13}/m^2_{F_1}$ and $\varepsilon^{u(d)}_{R23}=vm_{t(b)} Y^{u(d)}_{23}/m^2_{F_2}$, and ignore other $\varepsilon_{Rij}$ that  involve the light quark masses. We use the flavor mixing matrices of Eq.~(\ref{eq:VqLR}) to investigate the new flavor couplings of the Higgs and the weak gauge bosons below. 

 From the Yukawa couplings in Eq.~(\ref{eq:yukawa}), the SM Higgs couplings to the quarks in the flavor space are written as:
 \be
 -{\cal L}_{hqq} &= \bar q'_L V^q q'_R  h + H.c.\,,  \label{eq:hq'q'}
 \ed
 where $q'=(u,c,t,T_1, T_2)$ or $(d,s,b,B_1,B_2)$, $T_i$, and $B_i$ are the physical states of VLTQs and carry the electric charges of $2/3$ and $-1/3$, respectively, and $V^q$ is the mixing matrix for the $q$-type quark and is given by:
 \begin{align}								 
 V^q &= V^q_L \left( \begin{array}{cc}
							 			{\bf m}_{q} /v & {\bf Y^q} \\
										 0 & 0 \end{array} \right) V^{q\dagger}_R  \non \\
                                            & \approx 
											 \left( \begin{array}{cc}
								 			{\bf m}_{q} /v - {\bf Y}^q \pmb{ \varepsilon}^{q\dagger}_R & {\bf Y^q} \\
											 \pmb{ \varepsilon}^{q\dagger}_L  {\bf m}_q/v & \pmb{ \varepsilon}^{q\dagger}_L  {\bf Y}^q
											 \end{array} \right) \,. \label{eq:Vq}
 \end{align}
The small effects, such as ${\bf m}_{q} {\pmb \varepsilon}^q_R/v$, ${\pmb \varepsilon}^{q\dagger}_L {\bf Y}^q {\pmb \varepsilon}^{q\dagger}_R/v$, and ${\pmb \varepsilon}^{q\dagger}_L {\bf m}_q {\pmb \varepsilon}^q_R/v$,  have been dropped.  According to Eq.~(\ref{eq:Vq}),  the $h$-mediated FCNCs for the SM quarks (e.g., $h$-$q$-$q''$) are proportional to $m_{q,\, q''} v/m^2_F$. If the mass effects  of the first two generations of quarks are neglected, we have the flavor-changing Higgs interactions:
 \begin{align}
 -{\cal L}_{hqq''} &= - \bar u_{iL} C_{it} t_R h -  \bar d_{iL} C_{ib} b_R h +H.c.\,, \label{eq:hqq''}\\
 C_{it} &=  \frac{ m_t}{4v}  (   \zeta_{1i} \zeta_{13} + 2 \zeta_{2i} \zeta_{23} )\,,\non \\
 C_{ib} &=  \frac{ m_b}{4v}  ( 2  \zeta_{1i} \zeta_{13} + \zeta_{2i} \zeta_{23} )\,, \non 
 \end{align}
 where $u_{1(2)}=u(c)$ quark, $d_{1(2)}=d(s)$ quark, the definition of ${\bf Y}^q$ in Eq.~(\ref{eq:Yud}) is applied, and $\zeta_{ij} = v Y_{ij}/m_{F_i}$. The  $B_d-\bar B_d$ and $B_s-\bar B_s$ oscillations can be induced via the tree-level  Higgs mediation  in the VLTQ model.  Additionally, the BRs for the flavor-changing processes $t\to (c, u) h$, which are highly suppressed in the SM, become sizable. In addition to the new FCNC couplings, the flavor-conserving couplings are also modified:
  \be
  -{\cal L}_{hqq} = \frac{m_t}{v} \left[ 1 - \frac{\zeta^2_{13}+2 \zeta^2_{23} }{4}\right] \bar t_L t_R h +  \frac{m_b}{v} \left[ 1 - \frac{2 \zeta^2_{13}+\zeta^2_{23} }{4}\right] \bar b_L b_R h +H.c. \label{eq:hqq}
  \ed
The modification of Higgs coupling to the top quark will affect the SM Higgs production and decays in the $pp$ collisions at the LHC. If we take $m_{F_1}=m_{F_2}=1$ TeV, $Y_{ij} =1$, and $v=246$ GeV, the $h$ production cross section by the top-quark loop will be reduced by $9\%$ of the SM prediction. That is, the influence of new effects cannot be ignored arbitrarily. 
 
 From the flavor mixing matrix in Eq.~(\ref{eq:Vq}), we  can also obtain the SM Higgs interactions with the VLTQs as:
  \begin{align}
  -{\cal L}_{hFF} & = (\bar T_{1L}, \bar T_{2L})  \left( \begin{array}{cc}
                             \frac{ \xi_{11}}{4} & \frac{\xi_{12}}{2\sqrt{2}} \\
                              \frac{\xi_{21}}{2\sqrt{2}} & \frac{\xi_{22}}{2} \end{array} \right)
                                          \left(\begin{array}{c}
                                                T_{1R} \\
                                                T_{2R} \end{array} \right)\, h\,, \non \\
& + (\bar B_{1L}, \bar B_{2L})  \left( \begin{array}{cc}
                             \frac{ \xi_{11}}{2} & -\frac{\xi_{12}}{2\sqrt{2}} \\
                              -\frac{\xi_{21}}{2\sqrt{2}} & \frac{\xi_{22}}{4} \end{array} \right)
                                          \left(\begin{array}{c}
                                                B_{1R} \\
                                                B_{2R} \end{array} \right)\, h                 \label{eq:hFF}                        
  \end{align}
 with $\xi_{ij}  = \sum_k \zeta_{ik} Y_{jk} = m_{F_j}/v \sum_k \zeta_{ik} \zeta_{jk}$. Since VLTQs are color triplet states in $SU(3)_C$ and carry the same color charges as those of the SM quarks, the new couplings $hFF$ also contribute to the $h$ production cross section via the ggF channel. We will study their influence on the process  $pp\to h\to \gamma\gamma$ in the numerical analysis.

 \subsection{Weak interactions of SM quarks and VLTQs}
 
 By combining the  charged  weak interactions of the SM quarks with those of VLTQs in Eq.~(\ref{eq:VFF}), the charged current interactions of quarks can be formulated by:
 \be
 {\cal L}_{W} &=& -\frac{g}{\sqrt{2}} \bar {\bf u}_L \gamma^\mu V^L_{\rm CKM} {\bf d}_L W^+_\mu 
  -\frac{g}{\sqrt{2}} \bar {\bf u}_R \gamma^\mu V^R_{\rm CKM} {\bf d}_R W^+_\mu +H.c.\,,  \label{eq:Wud}
 \ed
where ${\bf u}=(u,c,t,T_1,T_2)$ and ${\bf d}=(d,s,b,B_1,B_2)$ are respectively the physical up- and down-type quarks, and $V^{L(R)}_{\rm CKM}$ is the $5\times 5$ CKM matrix for left (right)-handed quarks, defined by:
 \be
 V^L_{\rm CKM} &=& V^u_L \left(
\begin{array}{cc}
\left(V_{\rm CKM} \right)_{3\times 3} & {\mathbb 0}_{3\times 2}  \\ 
 {\mathbb 0}_{2\times 3} &  \sqrt{2} \mathbb{1}_{2\times 2}  
\end{array}
\right) %
 V^{d\dagger}_L \,, \ \  V^R_{\rm CKM} =  V^u_R 
  \left(
\begin{array}{cc}
{\mathbb 0}_{3\times 3} & {\mathbb 0}_{3\times 2}  \\ 
{\mathbb 0}_{2\times 3}&  \sqrt{2} \mathbb{1}_{2\times 2}  
\end{array}
\right)
 V^{d\dagger}_{R}\,.
 \ed
The $3\times 3$ matrix $V_{\rm CKM}$ is associated with the SM CKM matrix. Since the weak isospin of a triplet quark differs from that of a doublet quark, the new $5\times 5$ CKM matrices $V^{L,R}_{\rm CKM}$ are non-unitary. 
 
By using the results of Eq.~(\ref{eq:VqLR}),  the CKM matrix elements for the three-generation SM quarks are modified to be: 
 \be
 (V^{\rm SM}_{\rm CKM})_{i j} \to (V_{\rm CKM})_{i j}  + \frac{1}{2\sqrt{2}} \left( \zeta_{1i} \zeta_{1j} - \zeta_{2i} \zeta_{2j}\right)\,. \label{eq:MCKM}
 \ed
With $Y_{1i}, Y_{2i}<1$ and  $m_{F_{1,2}}=1$ TeV, the changes of the SM CKM matrix elements are roughly estimated as $(\Delta V^{\rm SM}_{\rm CKM})_{ij}= V^{\rm SM}_{\rm CKM} -V_{\rm CKM}  < 3\%$.  As indicated by experiments~\cite{PDG2014},  the value of $3\%$  has the same order of magnitude as  $(V_{\rm CKM})_{cb,ts}$  and is larger than $(V_{\rm CKM})_{ub,td}$. To satisfy the constraints of  $(V_{\rm CKM})_{ub,td}$, the possible schemes are: (a) $|\zeta^2_{ij}|$ is less than $|V_{ub}|$, the smallest  CKM matrix element, (b) $\zeta_{11,21}=0$ so that $(\Delta V^{\rm SM}_{\rm CKM})_{ub}=(\Delta V^{\rm SM}_{\rm CKM})_{td}=0$ and (c) $ \zeta_{11} \zeta_{13} - \zeta_{21} \zeta_{23}=0$ which  leads to $(\Delta V^{\rm SM}_{\rm CKM})_{ub}=(\Delta V^{\rm SM}_{\rm CKM})_{td}=0$. Moreover, if we adopt $\zeta_{1i}=\zeta_{2i}$ $(i=$1-3),  all CKM matrix elements return to the SM ones.  With the leading-order approximation for $\zeta_{ij}$, the $W$-boson interactions with the SM quarks and VLTQs are given by~\cite{Chen:2015cfa}:
 \begin{align}
 {\cal L}_{WFq} &=- \frac{g}{\sqrt{2} } \left[ -\frac{3 \zeta_{2i}}{2}  \bar u_{iL} \gamma^\mu B_{2L}  + \left( -\frac{\zeta_{1i}}{2} \bar T_{1L}  +\sqrt{2} \zeta_{2i} \bar T_{2L}  \right) \gamma^\mu d_{iL} \right]W^+_\mu   \non  \\
 &- \frac{g}{2} \left[ \zeta_{2i} \bar d_{iL} \gamma^\mu Y_L + \frac{m_b \zeta_{23}}{m_{F_2}} \bar b_R \gamma^\mu Y_R 
 - \zeta_{1i} \bar X_L \gamma^\mu u_{iL} - \frac{m_t \zeta_{13}}{m_{F_1}} \bar X_R \gamma^\mu t_R \right] W^+_\mu   
 +H.c.
 \label{eq:WFq}
 \end{align}
The charged weak interactions of VLTQs can be directly read from Eq.~(\ref{eq:VFF}). 

We next discuss the neutral weak interactions. It is known that  the left-handed and right-handed quarks in the SM are $ SU(2)_L$ doublets and singlets, respectively; however, the VLTQs are $SU(2)_L$ triplets. Since the isospin of a triplet is different from those of doublets and singlets, in order to combine the VLTQs with the SM quarks into  the same representation in the flavor space, we need to rewrite the vertex structure of the $Z$-boson, $T^3 - s^2_W Q$, in Eq.~(\ref{eq:VFF}) to fit the cases of doublets and singlets, such as $I_3 - s^2_W Q$, where $I_3 = \pm 1/2$ for doublets and $I_3=0$ for singlets.  Due to the isospin difference, $Z$-mediated FCNCs  are induced at the tree level. Since VLTQs $X$ and $Y$  carry the electric charges of $5/3$ and $ -4/3$, respectively,  they can not mix up with other quarks in the neutral current interactions. 
 
 In terms of weak eigenstates, we write the weak neutral current interactions in Eq.~(\ref{eq:VFF}) as:
  \begin{align}
  {\cal L}_{ZFF} &= - \frac{g}{c_W} Z_\mu \left[\bar {\cal F}_L \gamma^\mu (I_3 -s^2_W Q_{\cal F}) {\cal F}_L + \bar {\cal F}_R \gamma^\mu (-s^2_W Q_{\cal F}) {\cal F}_R   \right. \non \\
                         &\left. + \bar {\cal F}_L \gamma^\mu \left(\begin{array}{cc}
                                                    -1/2 & 0 \\
                                                    0 & 1/2 \end{array}\right) {\cal F}_L 
                            + \bar T_R \gamma^\mu \left(\begin{array}{cc}
                                                    0 & 0 \\
                                                    0 & 1 \end{array}\right) T_R 
                             + \bar B_R \gamma^\mu \left(\begin{array}{cc}
                                                    -1 & 0 \\
                                                    0 & 0 \end{array}\right) B_R \right]\,, \label{eq:ZFF}
  \end{align}
where $T^T= (U_1, U_2)$ and $B^T=(D_1, D_2)$ are composed of VLTQs with electric charges of $2/3$ and $-1/3$, respectively, ${\cal F}=T, B$, $I_3 = \pm 1/2$ for  $T (B)$, $Q_T = 2/3$, and $Q_B=-1/3$. We succeed in  expressing the $Z$ couplings to VLTQs by using the SM $Z$ couplings. It is clear that the first two terms in Eq.~(\ref{eq:ZFF}) lead to the flavor-conserving couplings when  the SM quarks and VLTQs form a representation in the dimension-5 flavor space. Since  the SM quarks do not have the interaction structures, as shown in the last three terms of Eq.~(\ref{eq:ZFF}), as a result, FCNCs via $Z$ mediation are generated.  Hence, the $Z$-boson interactions with  quarks, which carry electric charges of $2/3$ and $-1/3$, can be formulated as:
 \be
 {\cal L}_{Zq'q'} &=& -\frac{g}{c_W} C^{q'_L}_{ij} \bar q'_{iL} \gamma^\mu q'_{j L} Z_\mu - \frac{g}{c_W} C^{q'_R}_{ij} \bar q'_{i R} \gamma^\mu q'_{jR} Z_\mu\,, \\  \label{eq:Zq'q'}
  C^{q'_L}_{ij} &=&( I^{q'}_3 -s^2_W Q_{q'} )\delta_{ij} + \frac{1}{2} \left( -V^{q'}_{Li4} V^{q'*}_{Lj4} + V^{q'}_{Li5} V^{q'*}_{Lj5} \right)\,, \non \\
  C^{q'_R}_{ij} &=&-s^2_W Q_{q'} \delta_{ij} + \epsilon_{q'} (V^{q'}_{R})_{i \alpha_{q'}} (V^{q'*}_{R})_{\alpha_{q'} j}
  \ed
where $q'=(u,c,t,T_1,T_2)$ or $(d,s,b,B_1,B_2)$,  $V^{q'}_{L,R}$ are defined in Eq.~(\ref{eq:VqLR}),  $(\epsilon_u, \alpha_u)= (1, 5)$, and $(\epsilon_d, \alpha_d)= (-1, 4)$.

Using Eq.~(\ref{eq:VqLR}) and the leading-order approximation, the new gauge couplings of the $Z$-boson to  the SM quarks are given by:
 \be
 {\cal L}_{Zq_iq_j} = -\frac{g}{8c_W} ( a_q \zeta_{1i} \zeta_{1j} - b_q  \zeta_{2i} \zeta_{2j}) \bar q_{iL} \gamma^\mu   q_{jL} Z_\mu \,, \label{eq:Zqiqj}
 \ed
where $q_i$ denote the up- or down-type SM quarks,  $a_u=b_d=1$, and  $b_u=a_d=\sqrt{2}$. It can be seen that the FCNC effects can contribute to $\Delta F=2$ neutral meson mixings. A comparison  with the results in Eq.~(\ref{eq:MCKM}) indicates that the induced new coupling structures in charged and neutral currents are different. It is interesting to investigate the possible schemes that can simultaneously  satisfy the constraints from the CKM matrix elements and the data of neutral meson oscillations. The interactions of the $Z$-boson couple to one VLTQ and one SM quark are shown as:
 \begin{align}
 {\cal L}_{ZFq} & = -\frac{g}{4c_W} \bar u_{iL} \gamma^\mu \left( \zeta_{1i} T_{1L} - \sqrt{2} \zeta_{2i}T_{2L}\right) Z_\mu
  -\frac{g}{4c_W} \bar d_{iL} \gamma^\mu \left( \sqrt{2} \zeta_{1i}B_{1L}+  \zeta_{2i}B_{2L}\right)Z_\mu  \non \\
  & -\frac{g}{c_W} \left( -\frac{m_t \zeta_{23} }{\sqrt{2} m_{F_2}}\right) \bar t_{R} \gamma^\mu T_{2R}Z_\mu 
   -\frac{g}{c_W} \left( \frac{m_b \zeta_{13} }{\sqrt{2} m_{F_1}}\right) \bar b_{R} \gamma^\mu B_{1R}Z_\mu +H.c.
 \end{align}
One can get the $Z$ couplings to VLTQs from Eq.~(\ref{eq:VFF}). 

\section{Constraints and  top-quark FCNCs}

In this section, we discuss the constraints from low-energy $\Delta F=2$ processes and from the data of the SM Higgs production and decay into diphotons. 

\subsection{$P-\bar P$ mixings}
 
 From Eqs.~(\ref{eq:hqq''}) and (\ref{eq:Zqiqj}), we know that the $h$- and $Z$-mediated FCNCs appear and  contribute to the $\Delta F=2$ processes, such as $K-\bar K$ and $B_q-\bar B_q$ mixings, where the current experimental data can give strict constraints on the free parameters. Since the FCNC couplings in the up-type quarks are the same as those in the down-type quarks and the hadronic effects  in the $D$-meson system are dominated by unclear non-perturbative effects, we focus on $\Delta K=2$ and $\Delta B=2$ processes.  
 
  Following the notations in previous studies~\cite{Buras:2001ra,Buras:2012fs}, the transition matrix elements for $K-\bar K$ and $B_q-\bar B_q$ mixings are given by:
  \begin{align}
  M^K_{12}(h) & = - \frac{(\Delta^{bd}_L(h))^2}{2 m^2_h} \left[ C^{\rm SLL}_1 (\mu_h) \bar P^{SLL}_1(\mu_h) + C^{\rm SLL}_2 \bar P^{\rm SLL}_2(\mu_h) \right]\,, \\
  M^{K}_{12}(Z) &= \frac{(\Delta^{sd}_L(Z))^2}{2 m^2_Z} C^{\rm VLL}_1 (\mu_Z) \bar P^{\rm VLL}_1(K,\mu_Z)\,, \\
  M^{B_q}_{12}(Z) & = \frac{(\Delta^{bd}_L(Z))^2}{2 m^2_Z} C^{\rm VLL}_1 (\mu_Z) \bar P^{\rm VLL}_1(B_q, \mu_Z)\,.
  \end{align}
 $C^a_i$ is the Wilson coefficient with ${\cal O}(\alpha_s)$ QCD corrections, and $\bar P^a_i$ denotes the hadronic effects that include the renormalization group (RG) evolution from high energy to low energy, whose expressions are~\cite{Buras:2001ra,Buras:2012fs}:
 \begin{align}
 C^{\rm SLL}_1 &= 1 + \frac{\alpha_s}{4\pi} \left( -3 \ln\frac{m^2_h}{\mu^2_h } + \frac{9}{2}\right)\,,  ~ C^{\rm SLL}_2 =  \frac{\alpha_s}{4\pi} \left( -\frac{1}{12} \ln\frac{m^2_h}{\mu^2_h } + \frac{1}{8}\right)\,, \non \\
 C^{\rm VLL}_1 & = 1 + \frac{\alpha_s}{4\pi} \left( -2 \ln \frac{m^2_Z}{\mu^2_Z } + \frac{11}{3}\right)\,, ~ \bar P^{a}_i (P,\mu) = \frac{1}{3} m_P f^2_P P^{\rm a}_i (P, \mu_p)\,, \non \\
 P^{\rm SLL}_1(B_q,\mu_b) & =- \frac{5}{8} [\eta_{11}(\mu_b) ]_{\rm SLL} r_{B_q}  B^{\rm SLL}_{1}(\mu_b) -\frac{3}{2} [\eta_{21}(\mu_b)]_{\rm SLL} r_{B_q}  B^{\rm SLL}_{2}(\mu_b) \,, \non \\
 P^{\rm SLL}_2(B_q,\mu_b) & =- \frac{5}{8} [\eta_{12}(\mu_b) ]_{\rm SLL} r_{B_q}  B^{\rm SLL}_{1}(\mu_b) -\frac{3}{2} [\eta_{22}(\mu_b)]_{\rm SLL} r_{B_q}  B^{\rm SLL}_{2}(\mu_b) \,, 
 \end{align}
$P^{\rm VLL}_1(P, \mu_p) = \eta_{\rm VLL}(\mu_p) B^{\rm VLL}(P,\mu_p)$, where $m_P$ and $f_P$ are the mass and decay constant of the $P$-meson, respectively, $\mu_p= 2 (\mu_b)$ GeV for the $K (B_q)$-meson, $r_{B_q} = (m_{B_q}/(m_b + m_q))^2$, and the values of other hadronic effects and RG evolution effects are given in Table~\ref{tab:Values}. $\Delta^{q_i q_j}_L $ are  from the short-distance interactions of Eqs.~(\ref{eq:hqq''}) and (\ref{eq:Zqiqj}) and are written as:
 \begin{align}
 \Delta^{bq}_L(h) &=- \frac{m_b}{4v} \left( 2 \zeta_{13} \zeta_{1j} + \zeta_{23} \zeta_{2j} \right)\,, \non \\
 \Delta^{ds}_L(Z) & = \frac{g}{8 c_W} \left( \sqrt{2} \zeta_{12} \zeta_{11} - \zeta_{22} \zeta_{21} \right)\,, \non \\
 \Delta^{bq}_L(Z) & = \frac{8}{8 c_W} \left( \sqrt{2} \zeta_{13} \zeta_{1j} - \zeta_{23} \zeta_{2j} \right)\,.
 \end{align}
Since we have ignored the effects of light quark masses, the $h$-mediated FCNC has no contribution to $K-\bar K$ mixing. 

To constrain the parameters, we assume that the obtained $\Delta m_{P}$ in the model should be less than the experimental measurements. To understand the individual influences of  $h$ mediation and $Z$ mediation, we show their constraints separately. With $\Delta m_K=2 |Re M^K_{12}|$, $\Delta m_{B_q}=2 |M^{B_q}_{12}|$, and the inputs of Table~\ref{tab:Values}, we obtain the constraints as:

\noindent {$K-\bar K$ mixing :}
\begin{align}
 & |\sqrt{2} \zeta_{12} \zeta_{11} - \zeta_{22} \zeta_{21}| < 0.0013 ~ ( Z )\,. \label{eq:KC}
  \end{align}
 
\noindent {$B_d-\bar B_d$ mixing:}
 \begin{align}
& |2 \zeta_{13} \zeta_{11} + \zeta_{23} \zeta_{21}| <  0.053 ~ ( h )\,,  \non \\
& |\sqrt{2} \zeta_{13} \zeta_{11} - \zeta_{23} \zeta_{21}|< 0.0024 ~ ( Z )\,. \label{eq:BdC}
  \end{align}

\noindent {$B_s-\bar B_s$ mixing:}
 \begin{align}
 & |2 \zeta_{13} \zeta_{12} + \zeta_{23} \zeta_{22}| <  0.26\ ~ ( h )\,, \non \\
 & |\sqrt{2}\zeta_{13} \zeta_{12} -\zeta_{23} \zeta_{22}| < 0.012 ~ ( Z ).  \label{eq:BsC}
  \end{align}
 From these results, we find that the constraint from $\Delta m_K$ is only a factor of 2 stronger than that from $\Delta m_{B_d}$. Since the ratio $\sqrt{\Delta m_{B_d}/\Delta m_{B_s}}$ in experiments is very close to the Wolfenstein's parameter $\lambda \approx 0.22$~\cite{PDG2014}, the difference of a factor of $0.2$ between Eq.~(\ref{eq:BdC}) and Eq.~$(\ref{eq:BsC})$  is reasonable. 
   \begin{table}[!ht]
   \caption{ }
  \label{tab:Values}
 \begin{ruledtabular}
  \begin{tabular}{ccccc} 
  $m_K$[GeV] & $m_{B_d}$[GeV] & $m_{B_s}$[GeV] & $f_K$[GeV] & $f_{B_d}$[GeV] \\ \hline 
  0.497 & 5.28 & 5.37 & 0.16 & 0.186  \\ \hline
  $f_{B_s}$[GeV] & $\Delta m_K$[GeV] & $\Delta m_{B_d}$[GeV] & $\Delta m_{B_s}$[GeV] & $m_b$[GeV]\\ \hline 
   0.224 & $3.48 \times 10^{-15}$ & $3.37 \times 10^{-13}$ & $1.17 \times 10^{-11}$ & 4.8 \\ \hline
   $[\eta_{11}(\mu_b)]_{\rm SLL}$ & $[\eta_{12}(\mu_b)]_{\rm SLL}$ & $[\eta_{21}(\mu_b]_{\rm SLL}$ & $[\eta_{22}(\mu_b)]_{\rm SLL}$  & $[\eta(\mu_L)]_{\rm VLL}$\\ \hline 
   1.654 & 1.993 & -0.007 & 0.549 & 0.788  \\ \hline 
   $[\eta(\mu_b)]_{\rm VLL}$ & $B^{\rm VLL}_1 (K,\mu_L)$ & $B^{\rm VLL}_1 (B_q ,\mu_b)$ & $\alpha_s$ & $\mu_L$[GeV]\\ \hline
   0.842 & 0.57 & 1 & 0.118 & 2 
  \end{tabular}
  \end{ruledtabular}
\end{table}

 From the definition  $\zeta_{ij} = v Y_{ij}/m_{F_i}$, if we take $Y_{13,23}\approx 1$ and $m_{F_i}=1$ TeV, we have  $\zeta_{13,23}\approx 0.25$. It is interesting to understand whether the values of $\zeta_{11,21,12,22}$ could be the same orders of magnitude as $\zeta_{13, 23}\approx 0.25$ when the constraints of Eqs.~(\ref{eq:KC}), (\ref{eq:BdC}), and (\ref{eq:BsC}) are satisfied simultaneously. Recalling Eq.~(\ref{eq:MCKM}), in order to avoid the constraint from the CKM matrix elements, one of possible scheme is $\zeta_{1i}=\zeta_{2i}$. With this scheme, $Z$-mediated  $\Delta B_d$ will give the bound to be $\epsilon_{11, 21} < 0.013$. That is, it is difficult to require all values of  $\zeta_{ij}$ to be as large as $0.25$. To obey the constraints from CKM matrix elements and $\Delta F=2$ processes, one can adopt the modified scheme $\zeta_{11}=\zeta_{21}\ll 1$ and  $\zeta_{12(13)}=\zeta_{22(23)}$. As a result, the SM CKM matrix is not changed and $\Delta m_{K, B_d}$, via $Z$-mediated effects, can be automatically small; thus, the main constraint is from $\Delta m_{B_s}$. If we set $\zeta_{12(22)}\sim \zeta_{13(23)} = \epsilon$, from Eq.~(\ref{eq:BsC}), we get $\epsilon^2 < 0.087$ by $h$ mediation and $\epsilon^2 < 0.029$ (i.e., $\epsilon < 0.17$ ) by the $Z$-mediation. Clearly, the $Z$-boson FCNCs give a stronger bound on $\epsilon$.

\subsection{Constraint from diphoton Higgs decay}
 
 The Higgs measurement usually is described by the signal strength, defined as the ratio of observation to the SM prediction and expressed as:
  \be
  \mu_f = \frac{\sigma(pp\to h) BR(h\to f)}{\sigma(pp\to h)_{\rm SM} BR(h\to f)_{\rm SM}}\,, \label{eq:muf}
  \ed
 where $f$ stands for the possible channels. Although vector-boson fusion (VBF) can also produce the SM Higgs, we only consider the ggF process because it is the dominant one. Since the new flavor mixings directly affect the Higgs production and the Higgs decay to diphotons, we concentrate on the constraint from the diphoton channel (i.e., $f=\gamma \gamma$), where the current results measured by ATLAS and CMS are  $\mu_{\gamma\gamma}=1.17\pm 0.27$~\cite{Aad:2015gba}   and $\mu_{\gamma\gamma}=1.13\pm 0.24$~\cite{CMS}, respectively.

It is known that the $h$ production is dominated by the loop with a heavy quark; in the SM,  the top-quark loop gives the dominant contributions. Besides the top quark,  four heavy VLTQs, namely  $T_{1,2}$ and $B_{1,2}$, in the model can contribute to the Higgs production. In addition, they also affect the Higgs decay to diphotons. In order to understand their influence, we discuss the Higgs production and decay separately.  According to the couplings in Eq.~(\ref{eq:hqq}), the Higgs coupling to the top quark is modified as:
 \be
  \frac{m_t}{v} \to \frac{m_t}{v} \left( 1 - \frac{\zeta^2_{13} + 2 \zeta^2_{23}}{4}\right)\,.
 \ed
Therefore, the effective Lagrangian for $ggh$ by the top-quark loop can be obtained by multiplying the extra factor to the SM one, that is
 \be
 {\cal L}^t_{ggh} = \frac{\alpha_s}{16\pi v}  \left( 1 - \frac{\zeta^2_{13} + 2 \zeta^2_{23}}{4}\right) A_{1/2}(\tau_t)  h G^{a\mu \nu}G^a_{\mu \nu} \,,\label{eq:gght} 
 \ed
where $\tau_t=4 m^2_t/m^2_h$ and the loop function is
 \begin{align}
A_{1/2}(\tau) &=   -2 \tau [1+(1-\tau) f(\tau)^2]\,,  \label{eq:A12}\\
f(x) & =\sin^{-1}(1/\sqrt{x})\,. \non
  \end{align}
 Using the Higgs couplings to VLTQs in Eq.~(\ref{eq:hFF}),  
the effective Lagrangian for $ggh$ induced by the VLTQ loops can be formulated as:
  \begin{align}
{\cal L}^{\rm VLTQ}_{ggh} &= \frac{\alpha_s}{16\pi } \left( \sum_{i=1,2} \frac{3  \xi_{ii} }{4m_{F_i}} A_{1/2}(\tau_{F_i})  \right) h G^{a\mu \nu}G^a_{\mu \nu} \non \\
& \approx \frac{\alpha_s}{16\pi v} \frac{3 }{4 } \left[\sum_{i=1,2}(\zeta^2_{i2} + \zeta^2_{i3} )A_{1/2}(\tau_{F_i})  \right] h G^{a\mu \nu}G^a_{\mu \nu}\,,\label{eq:gghVL} 
 \end{align}
where $\tau_{F_i} = 4 m^2_{F_i}/m^2_h$ and  the small effects $\zeta^2_{11,21}\ll 1$ are dropped in the second line of the above equation. It is known that when  $\tau_{t, F_i} \to \infty$, $A_{1/2} \to -4/3$. The deviations of $A_{1/2}$ from the limit of $-4/3$ for $m_t=174$ GeV and $m_{F_i}=1$ TeV are only $3\%$ and $0.09\%$, respectively. Taking $A_{1/2}=-4/3$ as a good approximation, the effective interaction of $hgg$ that  combines Eq.~(\ref{eq:gght}) with Eq.~(\ref{eq:gghVL}) can be written as:
 \be
 {\cal L}_{ggh} =- \frac{\alpha_s}{12\pi v} \left[1 + \frac{1}{4} (3 \zeta^2_{12} + 2\zeta^2_{13} + 3 \zeta^2_{22} + \zeta^2_{23}) \right] h G^{a\mu \nu}G^a_{\mu \nu}\,.
 \ed 
If we adopt $\zeta_{12,13,22,23}\sim \epsilon $, the ratio of the Higgs production cross section to the SM result through the ggF process is easily obtained as:
 \be
 \frac{\sigma(pp\to h)}{\sigma(pp\to h)_{\rm SM}} \approx \left| 1 + \frac{9}{4} \epsilon^2    \right|^2\,.
 \ed
With $\epsilon=0.17$, the deviation from $1$ is around $13\%$. 

Next, we discuss the modification of the partial decay width for the decay $h\to \gamma\gamma$. Following the notations in a previous study~\cite{Gunion:1989we}, we write the partial decay width for $h\to \gamma\gamma$ as:
 \be
 \Gamma^{h}_{ \gamma\gamma}= \frac{\alpha m^3_h}{246 \pi^3 v^2} \left| \sum_{i} N_{ci} Q^2_i A_i (\tau_i) \right|^2\,,
 \ed
where $N_{ci}$ is the number of colors carried by the internal particle $i$, $Q^2_i$ is the electric charge square of particle $i$, and $A_i$ is the corresponding loop integral function. In the SM, the $W$-loop and the top-quark loop are the main effects. The loop function from the $W$-boson contribution is:
 \be
 A_{W}(\tau_W) = 2 + 3 \tau_W + 3\tau_W (2-\tau_W ) f(\tau_W)^2 \approx 8.34
 \ed
with $\tau_W =4 m^2_W/m^2_h$. The loop integral function from the top quark is $A_{1/2}$, which has been  defined in Eq.~(\ref{eq:A12}). Since the introduced VLTQs are spin-1/2 particles, the resulting loop integral function is also $A_{1/2}$ but with different argument $\tau_{F_i} = 4 m^2_{F_i}/m^2_h$. The modification of $\Gamma^h_{\gamma\gamma}$ can  thus be formulated as:
 \begin{align}
 \Gamma^h_{  \gamma\gamma} &= \Gamma^{h,\rm SM}_{ \gamma\gamma} \left| 1 + N_c \frac{ \zeta^t_{\gamma\gamma} + \zeta^{TB}_{\gamma\gamma}  }{A_W(\tau_W) + N_c Q^2_u A_{1/2}(\tau_t)} \right|^2 \,, \\
 \zeta^t_{\gamma\gamma} & = - \frac{Q^2_u }{4} \left( \zeta^2_{13} + \zeta^2_{23} \right) A_{1/2}(\tau_t) \,, \non \\
 \zeta^{TB}_{\gamma\gamma} & \approx \frac{Q^2_u + 2 Q^2_d }{4} \left( \zeta^2_{12} + \zeta^2_{13} \right) A_{1/2}(\tau_{F_1}) \non \\
 & +  \frac{2Q^2_u + Q^2_d}{4}    \left( \zeta^2_{22} + \zeta^2_{23} \right) A_{1/2}(\tau_{F_2}) \,, \non 
 \end{align}
 where $N_c=3$, $Q_u=2/3$, $Q_d=-1/3$, and the small effects $\zeta^2_{11,21} \ll 1$ in $\zeta^{TB}_{\gamma\gamma}$ have been neglected. As discussed earlier, it is a good approximation to use the limit $\tau_{t, F_i} \to \infty$, i.e., $A_{1/2} = -4/3$.  Using this limit and taking  $\epsilon_{12,13,22,23}\sim \epsilon$, the ratio of  $\Gamma^h_{\gamma\gamma}$ to the SM result can be simplified as:
 \be
 \frac{\Gamma(h\to \gamma\gamma)}{\Gamma(h\to \gamma\gamma)_{\rm SM}} \approx \left|1- N_c \frac{  (4Q^2_u/3 + 2 Q^2_d) \epsilon^2}{A_W (\tau_W) + N_c Q^2_u A_{1/2}(\tau_t)} \right|^2\,. \label{eq:Ghr}
 \ed
 With $\epsilon=0.17$, we find that the deviation of $\Gamma^h_{\gamma\gamma}$ from the SM result is only $-2\%$. 

 Since the influence of new physics on the Higgs width is small, the result in Eq.~(\ref{eq:Ghr}) can be regarded as the result of $BR(h\to \gamma\gamma)/BR(h\to \gamma\gamma)_{\rm SM}$. According to our analysis, if we take $\epsilon \lesssim 0.17$, the signal strength for diphoton Higgs decay defined in Eq.~(\ref{eq:muf}) is $\mu_{\gamma\gamma} \lesssim 10\%$.  This result is consistent with the current measurements at the LHC.

 \subsection{ Rare $t \to  q h$ and $t\to q Z$ decays} 
 
 It is known that the FCNCs in the SM arise from charged weak interactions through the loop effects. However, not all of them are sizable and detectable in the experiments, such as the rare top-quark decays $ t\to u_i h$ and $t\to u_i Z$ ($u_i=u,c$), in which the SM results are highly suppressed. As discussed earlier, the tree-level  $h$- and $Z$-mediated  FCNC couplings  to the SM quarks occur in this model. Following Eqs.~(\ref{eq:hqq''}) and (\ref{eq:Zqiqj}), the partial decay rates for $t \to u_i h$ and $t\to u_i Z$ are derived as:
 \begin{align}
 \Gamma(t\to u_i h) &= \frac{m_t }{32 \pi} |C_{it}|^2   \left( 1 - \frac{m^2_h}{m^2_t} \right)^2 \,, \\
 \Gamma(t\to u_i Z) &= \frac{m_t }{32 \pi} |C^Z_{it}|^2 \frac{m^2_t}{m^2_Z}\left( 1 +2 \frac{m^2_h}{m^2_t} \right)\left( 1 - \frac{m^2_h}{m^2_t} \right)^2\,, \\
 C^Z_{it} & = -\frac{g}{8 c_W}\left( \zeta_{1i} \zeta_{13} - \sqrt{2} \zeta_{2i} \zeta_{23} \right)\,.\non
 \end{align}
Taking $\zeta_{12}=\zeta_{22}$ and $\zeta_{13} =\zeta_{23}$, the constraints from  $\Delta F=2$ processes  in Eqs.~(\ref{eq:KC}), (\ref{eq:BdC}) and (\ref{eq:BsC}) can be directly applied. As a result, we get:
 \be
 BR(t\to (u, c) h) < (0.08, 6.8) \times 10^{-5}\,, \non \\
 BR(t \to (u, c) Z) < ( 0.19, 4.8)\times 10^{-6}\,. \label{eq:thZ}
 \ed
The current upper limits from ATLAS and CMS for $t\to c (u) h$ are $0.46(0.45)\%$~\cite{Aad:2015pja} and $0.47 (0.42)\%$~\cite{CMStqh} and for $t\to u_i Z$ are $7\times 10^{-4}$~\cite{Aad:2015uza} and $5 \times 10^{-4}$\cite{Chatrchyan:2013nwa}, respectively. It can be seen that  the results for the decays $t\to c (h,Z)$  in Eq.~(\ref{eq:thZ}) are smaller than  the current data by two orders of magnitude.

\section{Single Production of  $X_{\pm 5/3}$ and $Y_{\mp 4/3}$}

The introduced VLQs  in the model are $T_{1,2}$, $B_{1,2}$, $X$, and $Y$, where the first two VLQs  can be regarded as the partners of up- and down-type quarks that carry electric charges of $Q_u= 2/3$ and $Q_d =  1/3$, respectively; however, the exotic particles $X$ and $Y$  carry electric charges of $5/3$ and $-4/3$, respectively. Since   the couplings of $X$ and $Y$ to the SM particles are  QCD and charged weak interactions, in order to clearly understand  the production mechanism for the VLQs, we focus the studies on the VLQs $X$ and $Y$.  To present the production of VLQs and their antiparticles, we use the notations of $X_{\pm 5/3}$ and $Y_{\mp 4/3}$, where the subscript indicates the electric charge of the particle. 

The production cross section for a  VLQ pair is lower than that for a single VLQ when the mass of the VLQ is as heavy as 1 TeV; therefore, in this study  we discuss the single production of $X_{\pm 5/3}$ and $Y_{\mp 4/3}$ in detail. 
The relevant  free parameters  are the masses of VLQs and $\zeta_{ij}$. In the  numerical analysis, we adopt:
 \begin{align}
& m_X=m_Y=m_F \in [750, 1200] {\rm GeV};~ \zeta_{11}=\zeta_{21}=0.02;  \non \\
 & \zeta_{12}=\zeta_{13}=\zeta_{22}=\zeta_{23} =\zeta \in [0.1,0.3].
 \end{align}
These taken values are close to the constraints from the low-energy physics and from the Higgs measurements. We separately discuss the QCD and electroweak production processes below. To calculate the production  cross section in the $pp$ collisions at $\sqrt{s}=13$ TeV, we implement our model in CalcHEP~\cite{Belyaev:2012qa} and  adopt  {\tt CTEQ6L}  parton distribution functions (PDFs)~\cite{Nadolsky:2008zw}.

\subsection{QCD production channels }

Since  $X_{\pm 5/3}$ and $Y_{\mp 4/3}$ are color triplet fermions,  their couplings to the gluons are the same as those of the SM quarks. In this subsection, we discuss the VLQ production through the QCD processes. To compare  with the single production, we present the VLQ-pair production cross section with respect to $m_{F}$ in Table~\ref{tab:pair}, where  the QCD and electroweak effects are included and  $Q=X_{5/3}, Y_{-4/3}$. Since  QCD dominates the pair production, the $Q$-pair production cross section only depends on the mass of the VLQ. 

   \begin{table}[!ht]
   \caption{ Heavy quark pair production cross section in $pp$ collisions at $\sqrt{s}=13$ TeV, where   $Q=X_{5/3}, Y_{-4/3}$.}
  \label{tab:pair}
 \begin{ruledtabular}
  \begin{tabular}{cccccc} 
  $m_F$ [GeV] & 800 & 900 &  1000 & 1100 & 1200 \\ \hline                           
  $\sigma(pp\to Q \bar Q)$ [fb] &  88 & 42 & 22 & 11 & 06 \\
       \end{tabular}
  \end{ruledtabular}
\end{table}

\begin{figure}[hptb] 
\begin{center}
\includegraphics[width=4 in]{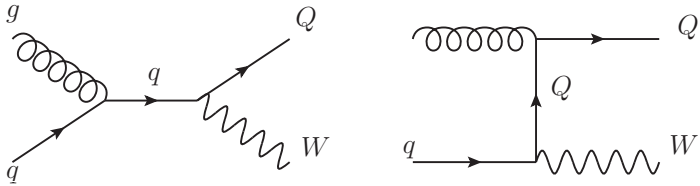} 
\caption{ Feynman diagrams for $gq\to Q W$, where $q$ is possible up (down) type-quarks while $Q=X_{5/3} (Y_{-4/3})$ and $W=W^{-} (W^+)$.  }
\label{fig:gq}
\end{center}
\end{figure}

As mentioned earlier, the relevant couplings of $X$ and Y to the SM particles are strong and charged weak interactions; therefore,  the production of a single VLQ in the final state via QCD effects is $gq \to Q W$, where  $q$ is the possible up (down)-type quarks while $Q=X_{5/3} (Y_{-4/3})$ and $W=W^{-} (W^+)$. The Feynman diagrams are sketched in Fig.~\ref{fig:gq}.  With $\zeta=0.2$ and $\zeta_{11,21}=0.02$, we show the production cross section for the $QW$ process with respect to $m_F$ in Table~\ref{tab:gqQW}. Since the values of CP-conjugate processes are close to the results in Table~\ref{tab:gqQW}, we do not show them repeatedly.
The $\zeta_{ij}$-dependence of the scattering amplitudes can be understood as follows:
\begin{equation}
M(q_i g \to X(Y) W) \propto \zeta_{1i} (\zeta_{2i}). 
\end{equation}

   \begin{table}[!ht]
   \caption{ Production cross sections for $X_{5/3}W^-$ and $Y_{-4/3}W^+$ with various values of $m_F$, where $\sqrt{s}=13$ TeV, $\zeta_{11,21}=0.02$, and $\zeta=0.2$ are used. }
  \label{tab:gqQW}
 \begin{ruledtabular}
  \begin{tabular}{cccccc} 
  $m_F$ [GeV] & 800 & 900 &  1000 & 1100 & 1200 \\ \hline                           
  $\sigma(pp\to X_{5/3} W^{-} )$ [fb] & 0.72 &  0.38 & 0.21 &  0.12 & 0.07 \\ \hline
  $\sigma(pp\to Y_{-4/3} W^+ )$ [fb] &  1.4 & 0.73 & 0.40 & 0.23 & 0.13  \\
       \end{tabular}
  \end{ruledtabular}
\end{table}

 From the results of Table~\ref{tab:gqQW}, it can be seen that except for the $Y_{-4/3} W^+$ mode with $m_{Y}=800$ GeV which  can have the cross section of ${\cal O}(1)$ fb, the others are below or far below 1 fb. In addition, $ \sigma(pp\to X_{5/3} W^{-} )$ is smaller than $\sigma(pp\to Y_{-4/3} W^+ )$ by a factor of two. These  results can be understood as follows. The quark $q$ in the $gq$ scattering is dominated by a sea quark, i.e., $c$-quark or $s$-quark. It is  known that the PDF of a sea quark is smaller than that of a valence quark when the momentum fraction is roughly larger than 0.1.  Therefore, the single production cross section of a VLQ via the $gq$ channel typically is small. Although the initial state for  $X$/$Y$ production can be the valence $u$/$d$ quark,  small production cross sections result from small couplings taken as $\zeta_{11,21}\ll 1$. If one uses $\zeta_{11,21}=1$ instead, the production cross section for $m_F =1$ TeV then can reach $75$ fb, which is a few factors larger than that for the case of pair production. For the same reasons, the  production cross sections for the CP-conjugate processes are the same as those shown in Table~\ref{tab:gqQW}. The factor of two difference between $X_{5/3} W^-$ and $Y_{-4/3} W^+$ actually arises from the different  PDFs in the initial quarks, where the averaged $c$-quark PDF for the former channel is a factor of around two smaller than the $s$-quark PDF for the latter channel.  In sum, we conclude that the single production cross section  of a VLQ via the $gq$ channel is far below 1 fb when the heavy quark mass approaches 1 TeV.

\subsection{Electroweak production channels}

One usually expects  that the production of a heavy quark is  dominated by  the strong interactions.  As we showed before, the heavy-quark-pair production cross section for $m_F=1$ TeV at $\sqrt{s}=13$ TeV is around 20 fb while the single production of a heavy quark is far below 1 fb.  In this subsection, we thoroughly  investigate the single production of $X$ and $Y$ through the electroweak interactions. We demonstrate that the single VLQ production cross section by electroweak interactions  is much larger than  that for  VLQ-pair production. Since the initial quarks for producing $X_{\pm 5/3}$ and $Y_{\mp 4/3}$ are different, we  discuss their situations  separately. 

\subsubsection{$X_{\pm 5/3}+$jet  processes}

$X_{5/3}$ accompanied by a quark jet can be  produced by $W$-mediated channels in $pp$ collisions, that is, $pp\to X_{5/3} q$, where $q$ can be anti-up-type quarks $\bar u_i=(\bar u, \bar c, \bar t)$ or down-type quarks $d_i=(d, s, b)$. Since the involved initial quarks  for $q=\bar u_i$ and $q=d_i$ final states are different, in order to understand the contributions from different situations, we  discuss them separately. Additionally, due to  the difference in PDF between the $u(d)$-quark and its anti-quark, we distinguish the CP-conjugate mode $X_{-5/3} \bar q$ from the $X_{5/3} q$ mode. 

\begin{figure}[hptb] 
\begin{center}
\includegraphics[width=4 in]{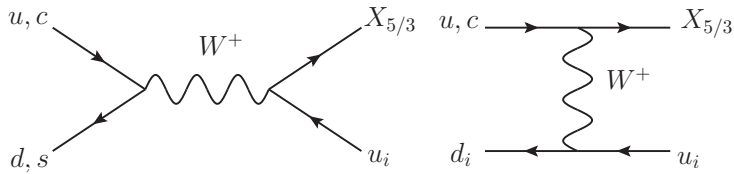} 
\caption{ $S$-channel (left) and $t$-channel (right) $q'\bar q''$ annihilation Feynman diagrams for production of $X_{5/3} \bar u_i$, where $\bar u_i = (\bar u, \bar c, \bar t)$.  }
\label{fig:xubar}
\end{center}
\end{figure}

We first study the $X_{5/3} \bar u_i$ processes. The possible Feynman diagrams are sketched in Fig.~\ref{fig:xubar}, where the left (right)-handed one is the $s(t)$-channel  $q' \bar q''$ annihilation diagrams.   
The $\zeta_{ij}$-dependence of the scattering  amplitudes is: 
\begin{equation}
M(u_i \bar d_j \to W \to X_{5/3} \bar u_k) \propto \zeta_{1k}, \quad M(u (c) d_i \to X_{5/3} u_i) \propto \zeta_{11} (\zeta_{12})\,,
\end{equation}
where the CP-conjugate processes have the same dependence. 
Since the off-diagonal CKM matrix elements are small, in the numerical estimations, we ignore their contributions.  Due to $\zeta_{11}\ll 1$, the processes that involve the vertex $u$-$X$-$W$ are small and their values are similar to  those shown in Table~\ref{tab:gqQW}. Since the coupling in the s-channel  $q' \bar q''$ annihilation to $X_{5/3} \bar u_i$ is the SM vertex $u$-$d$-$W$, unlike the case for the single VLQ production, the coupling  from the valence $u$-quark is not suppressed.

\begin{figure}[hptb] 
\begin{center}
\includegraphics[width=75mm]{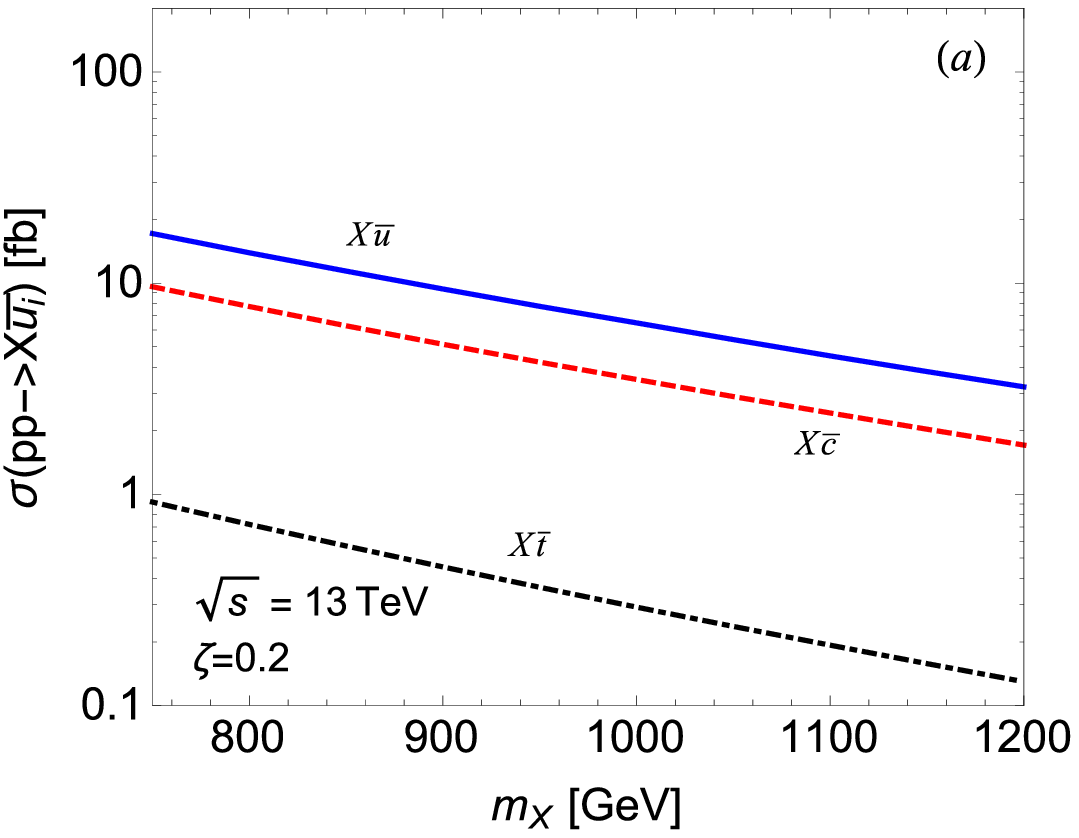} 
\includegraphics[width=75mm]{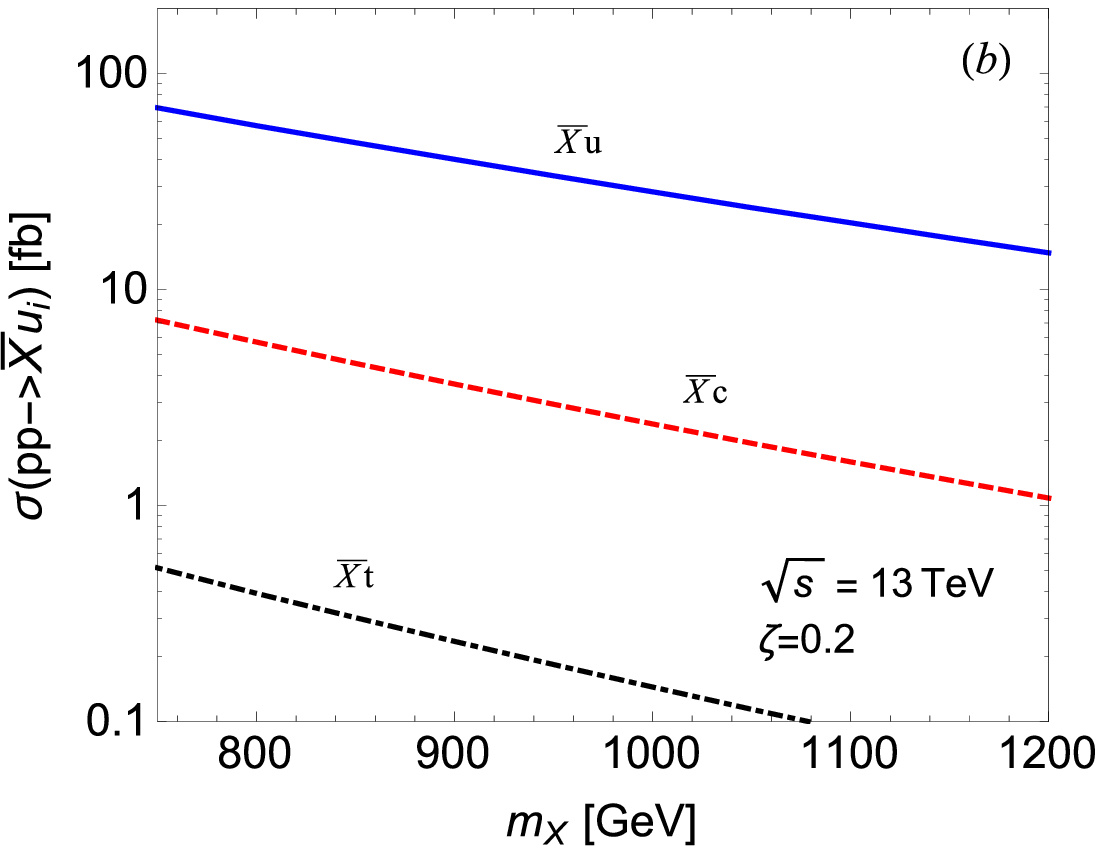} 
\caption{Production cross section ( in units of fb) as function of $m_X$ in $pp$ collisions at $\sqrt{s}=13$ TeV, where plot (a) is for $X_{5/3} (\bar u, \bar c, \bar t)$ processes while plot (b) is for $X_{-5/3} ( u, c, t)$. $\zeta_{11}=0.02$ and $\zeta=0.2$ are applied.  }
\label{fig:xu}
\end{center}
\end{figure}

We display the production cross sections for $X_{5/3} \bar u_i$ as a function of $m_X$ in Fig.~\ref{fig:xu}(a), where  $\sqrt{s}=13$ TeV, $\zeta_{11}=0.02$, and $\zeta=0.2$ are used.  It can be seen clearly that the relative magnitude of each production cross section is $\sigma(X_{5/3} \bar u) > \sigma(X_{5/3} \bar c) \gg \sigma(X_{5/3} \bar t)$. For $m_X=1$ TeV, we get $\sigma(X_{5/3} \bar u)=6.5$ fb, $\sigma(X_{5/3} \bar c)=3.5$ fb, and $\sigma(X_{5/3} \bar t)=0.3$ fb. We take $m_X=1$ TeV as the example to understand these results. The typical value of the cross section for $m_X=1$ TeV from the $s$-channel  $u\bar d \to X_{5/3} (\bar c, \bar t)$ is $0.1$ fb; however, it becomes $10^{-3}$ fb for $c \bar s \to X_{5/3}  \bar t$, where the suppressed cross section originates from the two sea quarks in the initial state. Accordingly, we can conclude that the production cross section that arises from the $s$-channel is far less than 1 fb. The results above 1 fb  are indeed  from the $t$-channel annihilations. For instance, the production cross section for the $t$-channel process $c \bar d \to X_{5/3} \bar u$ is $ 4.5 $ fb. We note that since $\bar b$ and $b$ have smaller PDFs, the cross section for $t$-channel $c \bar b \to X_{5/3} \bar t$ is of the order of $0.1$ fb. 

It is interesting to explore the difference between the CP-conjugate modes. With the same values of parameters, we present the $X_{-5/3} u_i$ production cross section via $W$ mediation as a function of $m_X$ in Fig.~\ref{fig:xu}(b). 
It is apparent that $\sigma(X_{-5/3} u)$ in $pp$ collisions is much larger than $\sigma(X_{5/3} \bar u)$, while the others are close to their CP-conjugate modes. For $m_X=1$ TeV, we have $\sigma(X_{-5/3} u)=28.3$ fb, $\sigma(X_{-5/3}c)=2.4$ fb, and $\sigma(X_{-5/3} t)=0.1$ fb. The  enhanced cross section for the $X_{-5/3} u$ mode originates from the valence $d$-quark, where  the associated PDF is larger, the process is dictated by  t-channel $ d \bar c  \to X_{-5/3} u$, and  the corresponding cross section is 27.6 fb. From the results,  we see clearly that the production cross section for the $X_{-5/3} u$ mode can be  as large as that  for VLQ-pair production.

\begin{figure}[hptb] 
\begin{center}
\includegraphics[width=75mm]{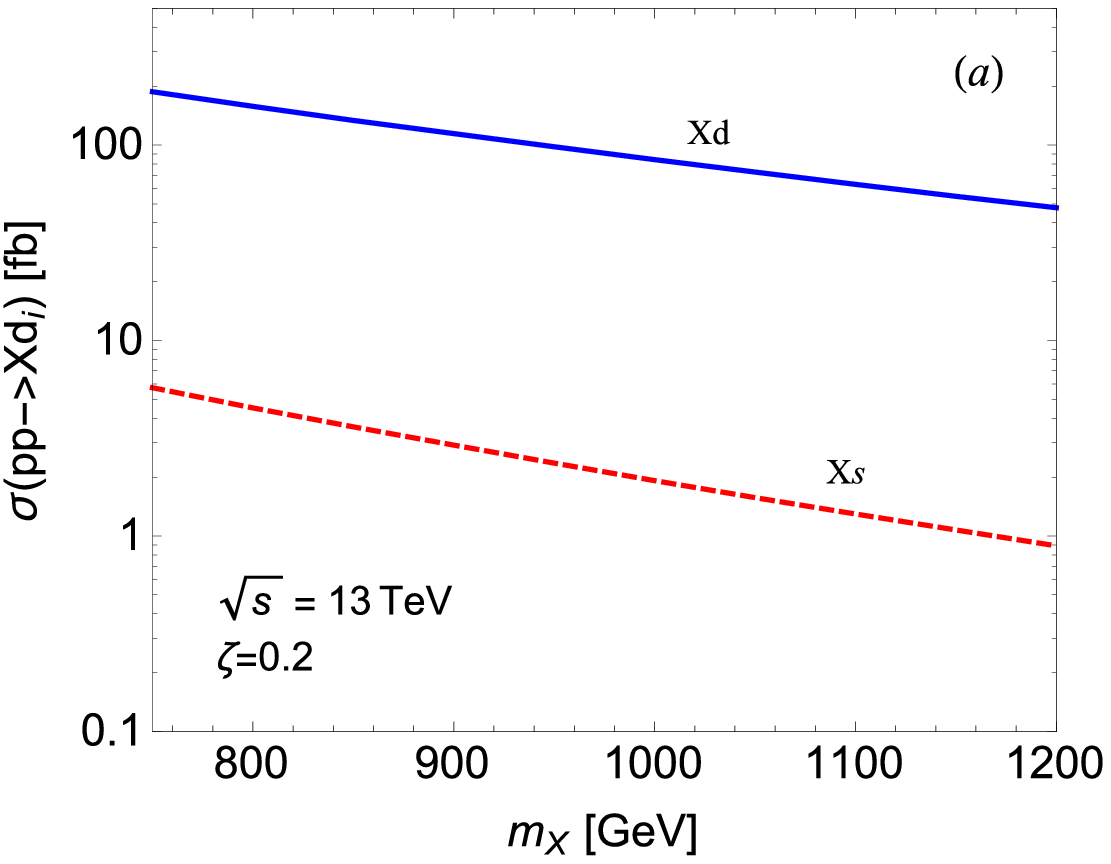} 
\includegraphics[width=75mm]{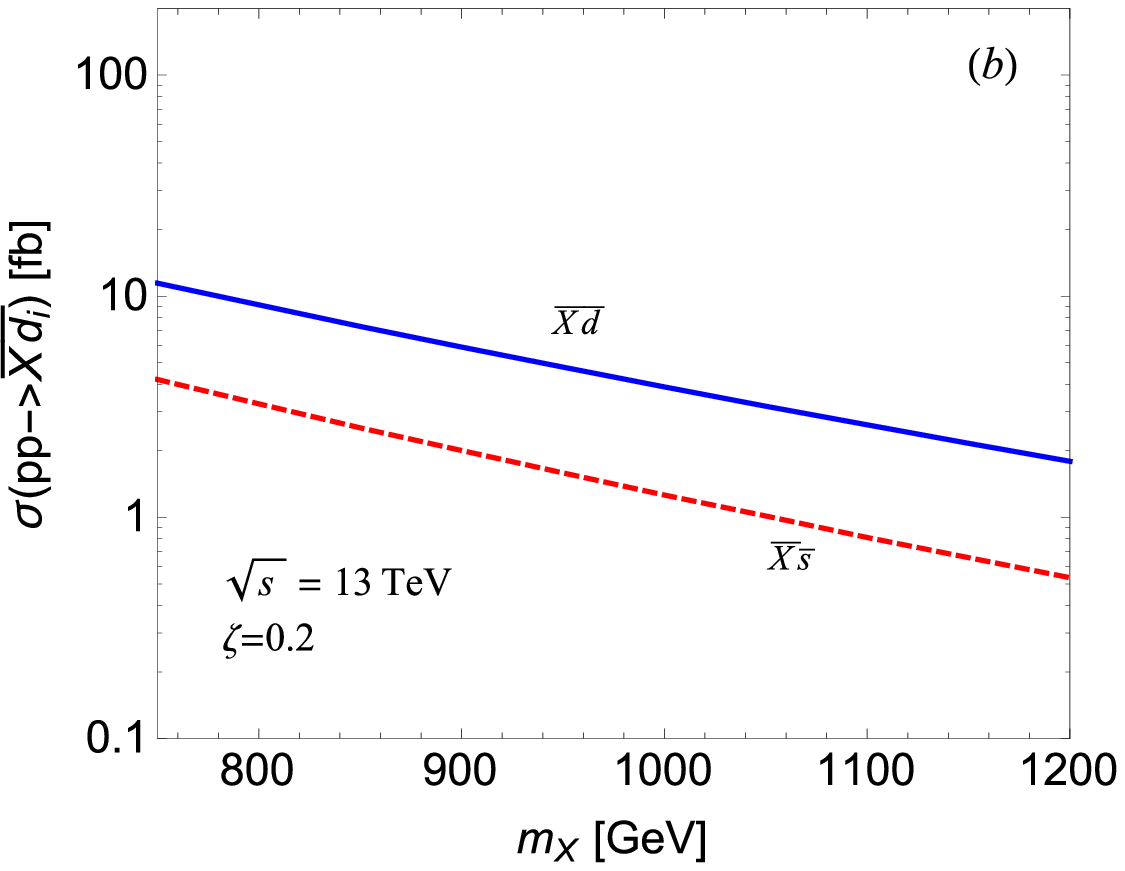} 
\caption{Production cross section ( in units of fb) as  function of $m_X$ in $pp$ collisions at $\sqrt{s}=13$ TeV, where plot (a) is for $X_{5/3} (d, s)$ processes while plot (b) is for $X_{-5/3} ( \bar d, \bar s)$. $\zeta_{11}=0.02$ and $\zeta=0.2$ are applied.  }
\label{fig:xd}
\end{center}
\end{figure}

In addition to $X_{5/3}\bar u_i$ and $X_{-5/3} u_i$ modes, where the net electric charges in the final state are $\pm 1$, we find that  $X_{5/3} d_i$ and $X_{-5/3}\bar d_i$ ( $d_i=d,s$) modes, whose electrical charge is $\pm 4/3$, are allowed and important. Since the net charges of the initial quarks have to be $\pm 4/3$, the possible combinations of quarks  are $uu$, $uc$, $cc$, and their antiparticles.  Therefore, only $t$-channel annihilation diagrams are involved. If the initial quarks are composed of  $\bar u_i \bar u_j$, since they are sea quarks, we expect that the resulting cross sections to be  similar to those  for $X_{5/3} \bar u_i$.  However, the situations for $u_i u_j$ are different. For instance, the processes for $X_{5/3} d$ can be classified as (i) $u u \to X_{5/3} d$ and (ii) $c u \to X_{5/3} d$, 
where the $\zeta_{ij}$-dependence of the scattering amplitudes is given by:
\begin{equation}
M(uu \to X_{5/3} d) \propto \zeta_{11}\, , \quad M(c u \to X_{5/3} d) \propto \zeta_{12} \,.
\end{equation}
Although process (i) depends on the small coupling $\zeta_{11}$,  the two large $u$-quark  PDFs compensate the suppression. For process (ii), although it involves a sea quark $c$ in the initial state, the related coupling is $\zeta$ and one $u$-quark PDF can enhance the contributions. With $m_X=1$ TeV and $\zeta=0.2$, we get $\sigma(uu\to X_{5/3} d) =14.5$ fb and $\sigma(uc\to X_{5/3} d) =69.8$ fb. Clearly, the production cross section for the $X_{5/3} d$ mode can be over 80 fb. Since the situation of the $X_{5/3} s$ mode is similar to that of the $X_{5/3} c$ mode, we expect that its production cross section is of the order of a few fb.  We present the  production cross sections for $X_{5/3} d_i$ and $X_{-5/3}\bar d_i$ modes as a function of $m_X$ in Figs.~\ref{fig:xd}(a) and (b), respectively. It can be seen that the $X_{5/3} d$ production cross section can be over 100 fb if the mass of the VLQ is lighter than 950 GeV. Obviously,  this result is higher than that for  VLQ-pair production.

\subsubsection{$Y_{\mp 4/3}+$jet processes }

 We discuss the single production of $Y_{-4/3}$ in this subsection. Similar to the production of $X_{\pm 5/3}$,  $Y_{-4/3}$  accompanied by a quark jet can be generated through the $W$-mediated processes and is described by $pp\to Y_{-4/3} \bar q$, where $\bar q= \bar d_i=(\bar d, \bar s, \bar b)$ or $\bar q = u_i = (u, c, t)$. Since the involved PDFs for the CP-conjugate modes are different, we discuss $Y_{-4/3} \bar q$ and $Y_{4/3} q$ modes separately.  

\begin{figure}[hptb] 
\begin{center}
\includegraphics[width=4 in]{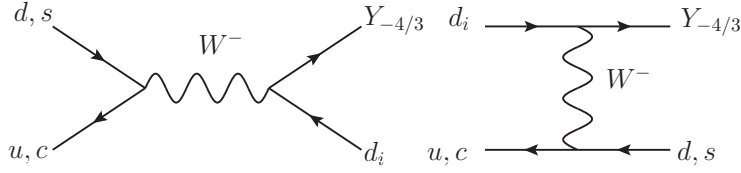} 
\caption{ $S$-channel (left) and $t$-channel (right) $d_i \bar u_j$ annihilation Feynman diagrams for production of $Y_{-4/3} \bar d_i$, where $\bar d_i = (\bar d, \bar s, \bar b)$.  }
\label{fig:ydbar}
\end{center}
\end{figure}

\begin{figure}[hptb] 
\begin{center}
\includegraphics[width=75mm]{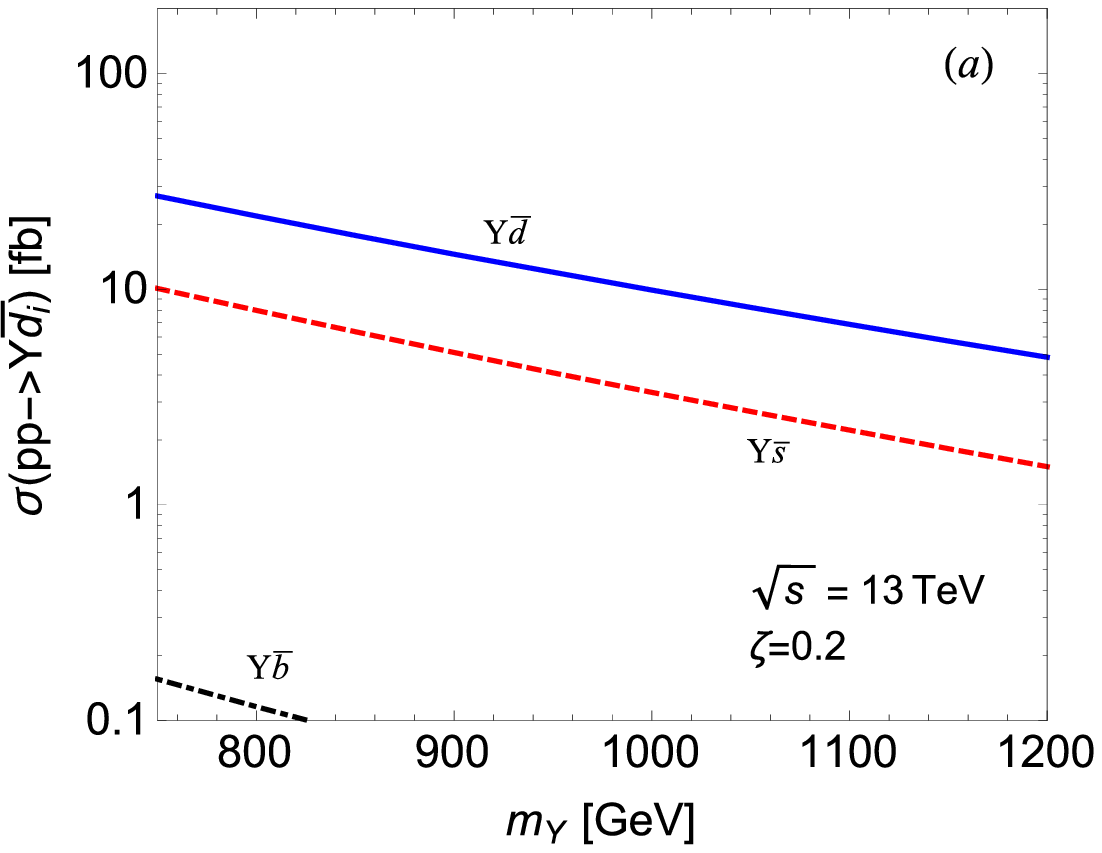} 
\includegraphics[width=75mm]{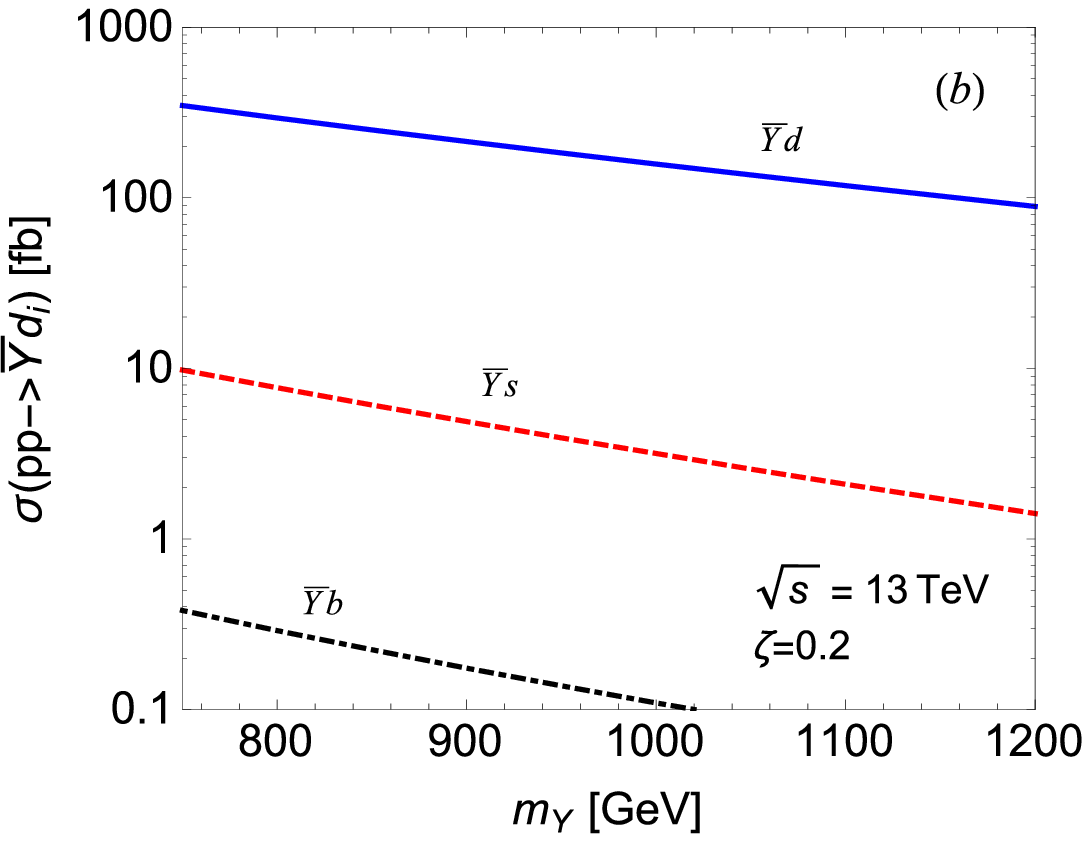} 
\caption{Production cross section ( in units of fb) as  function of $m_Y$ in $pp$ collisions at $\sqrt{s}=13$ TeV, where plot (a) is for $Y_{-4/3} \bar d_i$ processes while plot (b) is for $Y_{4/3}  d_i$, where $\zeta_{21}=0.02$ and $\zeta=0.2$ are applied.  }
\label{fig:yd}
\end{center}
\end{figure}

We first discuss the $Y_{-4/3} \bar d_i$ final states, in which  the $W$-mediated processes are through the $s$- and $t$-channel $d_i \bar u_j$ annihilations, and  the corresponding Feynman diagrams are shown in Fig.~\ref{fig:ydbar}. 
The $\zeta_{ij}$-dependence of the scattering amplitudes is read  as:
\begin{equation}
M(\bar u_i d_j \to W \to Y_{-4/3} \bar d_k) \propto \zeta_{2k}\,, \quad M(\bar u (\bar c) d_i \to Y_{-4/3} \bar d( \bar s )) \propto \zeta_{2i}\,,
\end{equation}
where the same dependence can be applied to their CP-conjugate processes. 
Based on the analysis for the single $X_{\pm 5/3}$ production, one expects $t$-channel annihilation diagrams to be dominant. With $m_Y=1$ TeV, $\zeta_{21}=0.02$, and $\zeta=0.2$, we illustrate the production cross section for the $s$-channel  to be $\sigma(d\bar u \to Y_{-4/3} (\bar s, \bar b) )= 3.8 \times 10^{-2}$ fb; however, it becomes $\sigma(s\bar c \to Y_{-4/3}  \bar b ) = 1.4 \times 10^{-3}$ fb when the initial state involves two sea quarks. Since the net electric charges of $Y_{-4/3} \bar d_i$ processes are $-1$  and  their initial states mostly involve two sea quarks, the $t$-channel production cross sections for $(s, b) \bar u \to Y_{-4/3} \bar d$ and $s \bar c \to Y_{-4/3} \bar s$ are $( 6.9,  2.3)$ fb. Although the $t$-channel $d (\bar u, \bar c) \to Y_{-4/3} (\bar d, \bar s)$ processes have one valence $d$-quark in the initial state, due to the small coupling $\zeta_{21}\ll 1$, their cross sections are suppressed to be $(0.68, 0.27)$ fb.  We present the production cross sections for $Y_{-4/3} \bar d_i$ modes as a function of $m_Y$  in Fig.~\ref{fig:yd}(a), where $\sqrt{s}=13$ TeV, $\zeta_{21}=0.02$, and $\zeta=0.2$ are used. For $m_Y=1$ TeV, we have $\sigma(Y_{-4/3} \bar d)=10$ fb, $\sigma(Y_{-4/3} \bar s)= 3.3$ fb, and  $\sigma(Y_{-4/3} \bar b)=0.04$ fb. All of these results can be understood from  the  discussions for $pp\to X_{5/3} \bar u_i$. 

It is interesting to examine the processes $pp\to Y_{4/3} d_i$ which  is the  CP-conjugate modes of $Y_{-4/3}\bar d_i$. As mentioned earlier,  the $t$-channel $\sigma(s \bar u \to Y_{-4/3} d)$ with sea quarks in the initial state can be of the order of $10$ fb; since the CP-conjugate process is $u \bar s \to Y_{4/3} \bar d$, where the initial state involves a valence $u$-quark, the cross section for such mode should be much larger than $Y_{-4/3} d$. With $m_Y=1$ TeV, $\zeta_{21}=0.02$, and $\zeta=0.2$, we obtain $\sigma(u\bar s \to Y_{4/3} d)= 111$ fb and $\sigma(u \bar b \to Y_{4/3} d)=45$ fb. The production cross section for the $Y_{4/3} d$ mode is one order of magnitude larger than that for the $Y_{-4/3} \bar d$ mode. Since the production mechanism  for $Y_{4/3} (s, b)$ modes is similar to that for  their CP-conjugate modes,  it is expected that the results are close to $Y_{-4/3} (\bar s, \bar b)$. In order to clearly see the numerical results, we plot the production cross sections for $Y_{4/3} d_i$ as a function of $m_Y$ in Fig.~\ref{fig:yd}(b), where $\sqrt{s}=13$ TeV, $\zeta_{21}=0.02$, and $\zeta=0.2$ are applied.

\begin{figure}[hptb] 
\begin{center}
\includegraphics[width=75mm]{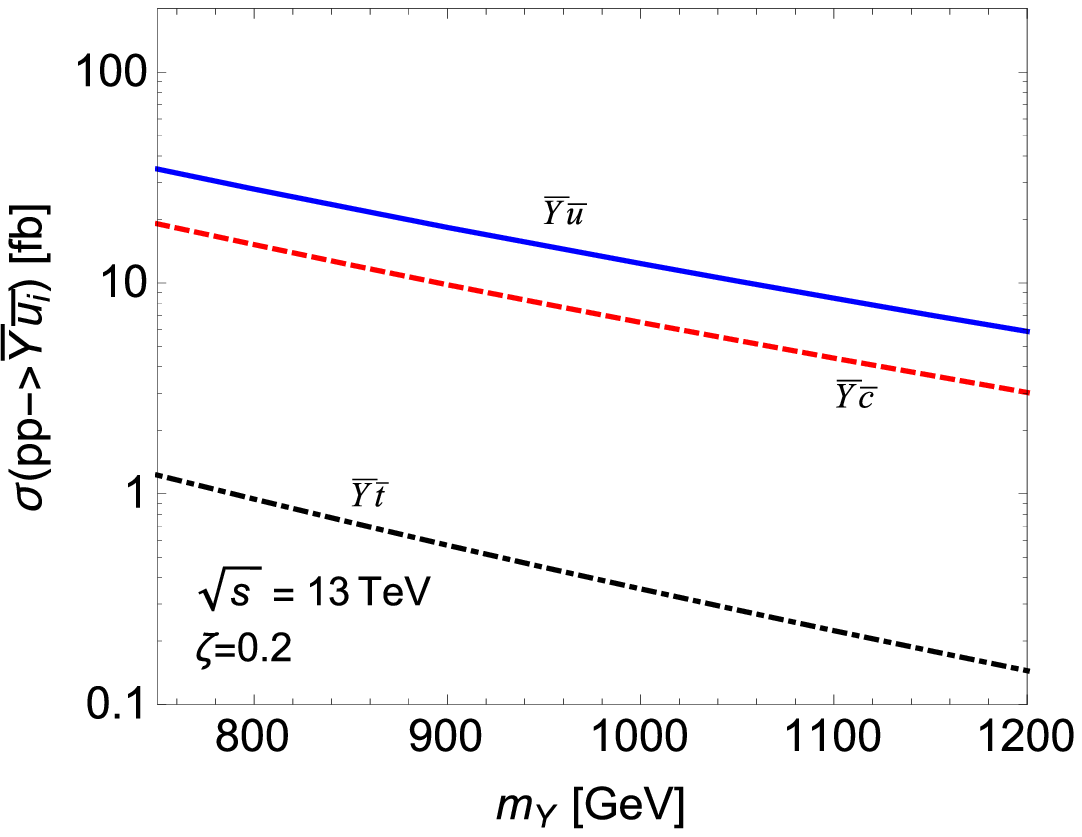} 
\includegraphics[width=75mm]{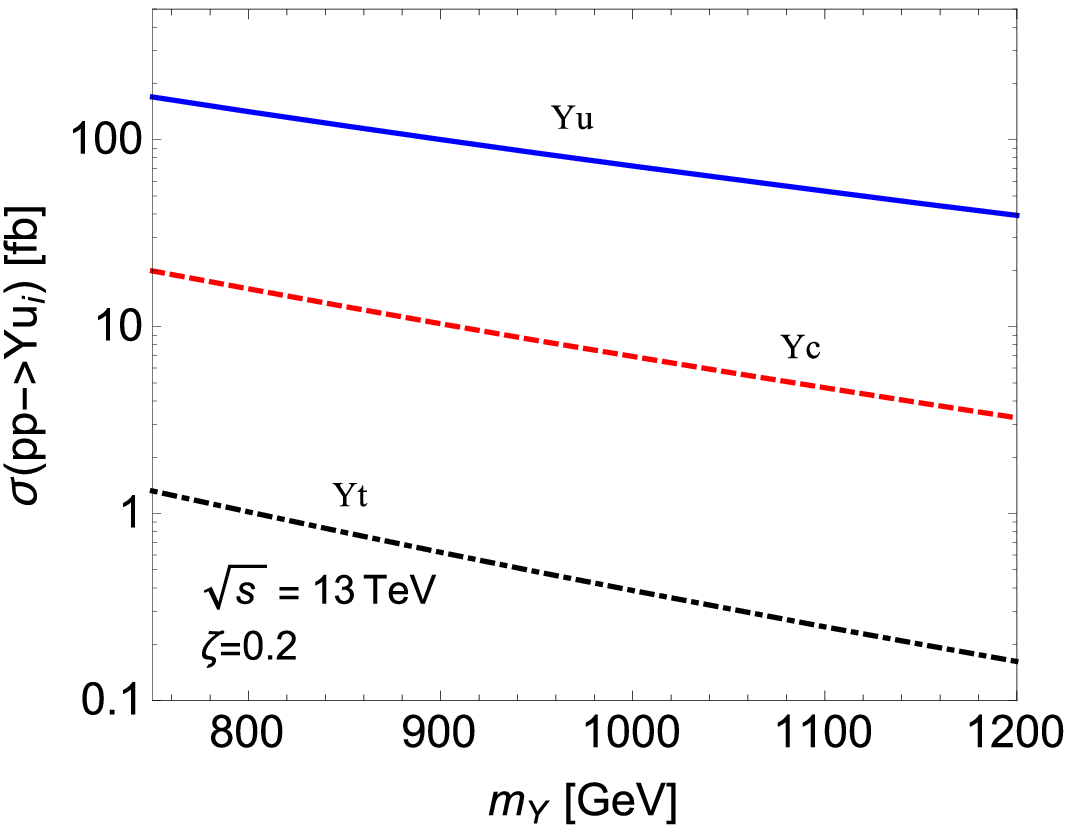} 
\caption{Production cross section ( in units of fb) as  function of $m_Y$ in $pp$ collisions at $\sqrt{s}=13$ TeV, where plot (a) is for $Y_{4/3} \bar u_i$ processes while plot (b) is for $Y_{-4/3} u_i$, where $\zeta_{21}=0.02$ and $\zeta=0.2$ are applied.  }
\label{fig:yu}
\end{center}
\end{figure}

Besides $Y_{-4/3} \bar d_i$ and $Y_{4/3} d_i$, in which  the net charges of final states are $\mp 1$, we can have the single $Y_{\mp 4/3}$ associated with a $u_i/\bar u_i$ quark in the final state, in which the net charges are $\mp 2/3$ and the $\zeta_{ij}$-dependence of the scattering amplitudes including their CP-conjugate processes is: 
\begin{equation}
M(d_i d_j \to Y_{-4/3} u_j ) \propto \zeta_{2i}\,. 
\end{equation}
The single $Y_{\mp 4/3}$ production channels are thus from $d_i d_j$ and $\bar d_i \bar d_j$ scatterings. Since the initial states $\bar d_i \bar d_j$  are the sea quarks,  it can be expected  that the resultant production cross sections should be similar to those of the processes $pp\to Y_{-4/3} \bar d_i$.   The production channels from $d_i d_j$ scatterings can have larger cross sections. 
For instance, with $m_Y=1$ TeV and $\zeta=0.2$, we get $\sigma(s d \to Y_{-4/3} u)=49$ fb and $\sigma(b d \to Y_{-3/4} u)=19$ fb. In addition, even though the process $d d \to Y_{-4/3} u$ involves the coupling $\zeta_{21}=0.02$, its contribution can still reach $\sigma(d d \to Y_{-4/3} u)=3.6$ fb. As to the $Y_{-4/3} c$ mode, its result is similar to that of $Y_{4/3} s$. We should mention that unlike the $Y_{4/3} b$ and $Y_{-4/3} \bar b$ modes, which are from the $s$-channel, $Y_{-4/3} t$ and $Y_{4/3} \bar t$ are mainly from the $t$-channel $s b$ and $\bar s \bar b$ annihilations, respectively. Although the cross sections are still small, they are larger than those for $Y_{4/3} b$ and $Y_{-4/3} \bar b$ modes. In sum, we have $\sigma(pp \to Y_{-4/3} u) = 72$ fb, $\sigma(pp \to Y_{-4/3} c)=6.9$ fb, and $\sigma(pp \to Y_{-4/3} t)=0.4$ fb. We numerically present the production cross sections for $Y_{4/3} \bar u_i$ and $Y_{4/3} u_i$ modes as a function of $m_Y$ in Fig.~\ref{fig:yu}, where $\sqrt{s}=13$ TeV, $\zeta_{21}=0.02$, and $\zeta=0.2$ are applied. 

\begin{figure}[t] 
\begin{center}
\includegraphics[width=75mm]{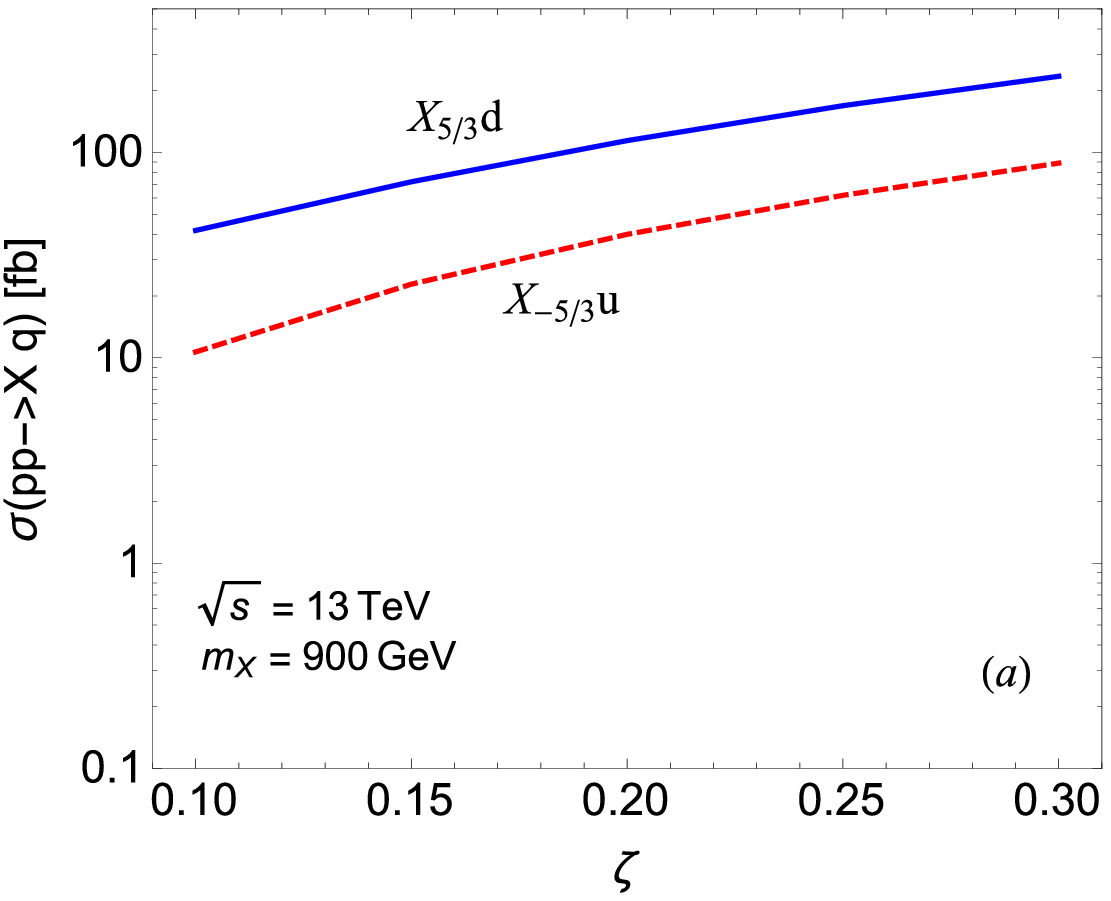} 
\includegraphics[width=75mm]{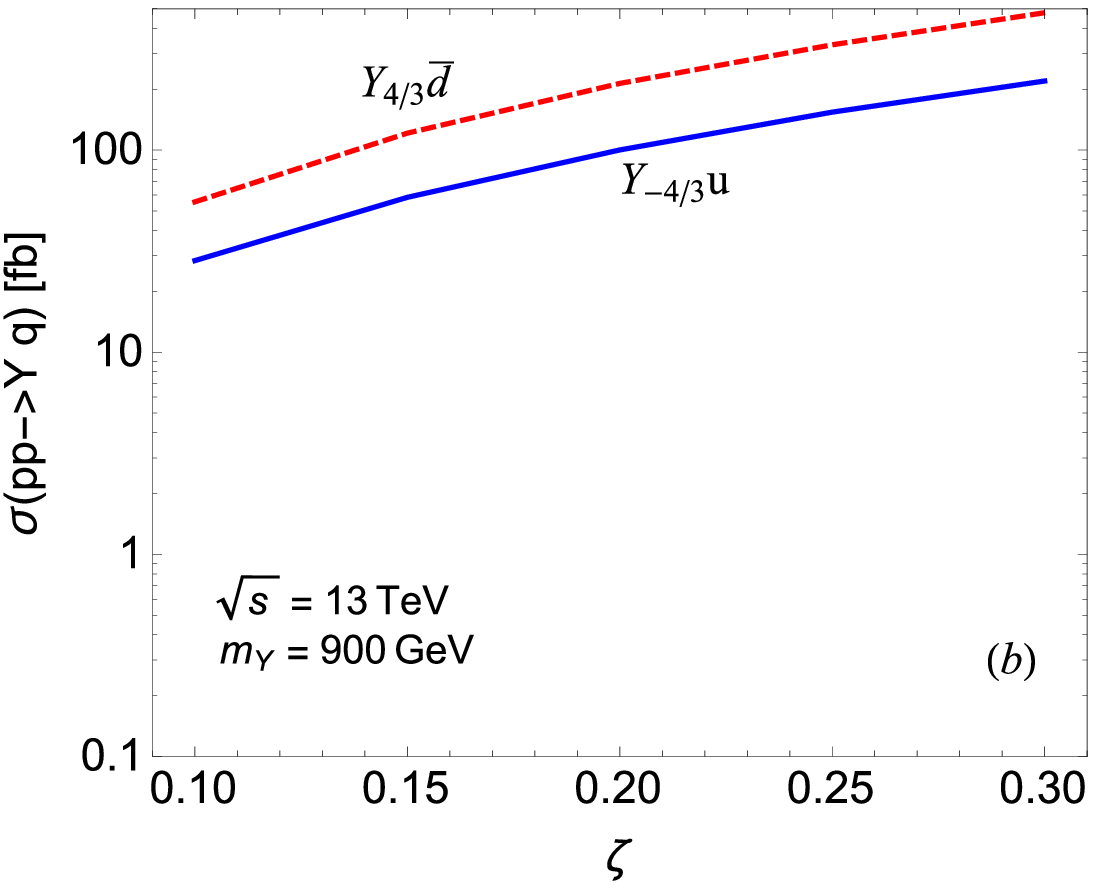} 
\caption{Production cross section ( in units of fb) as  function of $\zeta$ in $pp$ collisions at $\sqrt{s}=13$ TeV, where plot (a) is for $X_{5/3} d$ (solid) and $X_{-5/3} u$ (dashed) processes while plot (b) is for $Y_{-4/3} u$ (solid) and $Y_{4/3} d$ processes, where mass of VLQ is set to be 900 GeV.  }
\label{fig:xy}
\end{center}
\end{figure}

To further understand the $\zeta$ dependence of  the single $X$ and $Y$ production processes, we show the production cross section as a function of $\zeta$ in Fig.~\ref{fig:xy}, where we use $m_X=m_Y=900$ GeV. Plot (a) is for $X_{5/3} d$ (solid line) and $X_{-5/3} u$ (dashed line), and the plot (b) is for $Y_{-4/3} u$ (solid line) and $Y_{4/3} d$ (dashed line). With the scheme $\zeta_{12}\approx \zeta_{13}\approx \zeta_{22} \approx \zeta_{23}$, the main decay channels for $X_{5/3}$ and $Y_{4/3}$ are $X_{5/3}\to W^+ (t, c)$ and $ Y_{4/3}\to W^+ (\bar s, \bar b)$, respectively,  and each BR is almost equal to 1/2. 
 Hence, the favorable channels to search for the single production of VLQs $X$ and $Y$ are
  \begin{align}
  \label{eq:signal}
  pp & \to d X_{5/3} \to d W^+ c \,,\non \\
  pp &\to d X_{5/3} \to d W^+ t \to d W^+ (W^+ b)\,, \non \\
  pp & \to d Y_{4/3} \to d W^+ ( \bar s, \bar b)\,.
  \end{align}
  
In the following we briefly simulate  the signals for the proposed processes.  Since the final states involve $W$-boson, we focus on the leptonic decays of $W$-boson. Thus, the signal events for $d W^+ c (\bar s, \bar b)$
 and $dW^+ t$ are $\ell^+ + {\rm jets}$  and $\ell^+ \ell^+ + {\rm jets}$ respectively, where $\ell=e, \mu$ and the number of jets is set to be $n_{\rm jet} \geq 2$. As to the backgrounds from the SM contributions,  we consider the processes $pp \to W^+(Z) j j$  and $pp \to W^+W^+(ZZ) j j$. To generate the signal and background events, we use the event generator {\tt MADGRAPH/MADEVENT\,5}~\cite{Alwall:2014hca}, 
where we have employed FeynRules 2.0 \cite{Alloul:2013bka} to create the relevant  Feynman rules and  parameters of the model, and we apply the {\tt NNPDF23LO1} PDF~\cite{Deans:2013mha}. 
We use {\tt PYTHIA\,6}~\cite{Ref:Pythia} to include hadronization effects,  the  initial-state radiation (ISR) and final-state radiation (FSR) effects, and the decays of the SM particles. 
Additionally, the generated events are run through the {\tt PGS\,4} to perform detector level simulation~\cite{Ref:PGS}.

In order to reduce the backgrounds, we adopt the following kinematical cuts:
\begin{align}
\label{eq:cut}
& p_T(j_{\rm leading}) > 300 \, {\rm GeV}, \quad p_T(j) > 30 \, {\rm GeV}, \quad p_T(\ell) > 30 \, {\rm GeV} \qquad  {\rm for\  } \ell^+ + {\rm jets}, \nonumber \\
& p_T(j_{\rm leading}) > 100 \, {\rm GeV}, \quad p_T(j) > 30 \, {\rm GeV}, \quad p_T(\ell) > 30 \, {\rm GeV} \qquad {\rm for\ } \ell^+ \ell^+ + {\rm jets}.
\end{align}
With the  cuts, we present the numbers of signal ($S$) and background ($B$) events and the significance $S/\sqrt{B}$ with a luminosity of 100 fb$^{-1}$ in Table~\ref{tab:event}, where $m_F=900$ GeV and $\zeta=0.1, 0.2$ are used. 
It is found that the significances of the channels $d W^+ c$ and $ d W^+ (\bar s, \bar b) $ are small; however,  since the processes with two same-sign leptons in the final state have smaller backgrounds, therefore, the channel $d W^+ t$ has a larger significance.
We believe that the significance can be further  improved by imposing more strict kinematical cuts.
The detailed event simulations will be studied in another paper. 
\begin{table}[hptb]
\caption{ Numbers of signal and background (BG) events and the significance of signal,  where we adopt a luminosity of  100 fb$^{-1}$, $m_F=900$ GeV, and $\zeta=0.1, 0.2$,  and  the kinematic cuts shown in Eq.~(\ref{eq:cut}) are applied.}
\label{tab:event}
\begin{ruledtabular}
\begin{tabular}{c||cc|cc} 
BG  \quad &  $W^+ j j$ \quad & $Z j j $  \quad & $W^+ W^+ j j$ \quad & $Z Z j j$ \quad  \\ \hline
 & $5.17 \times 10^5$ & $8.93 \times 10^4$   & 161  &  94.1
 \\ \hline \hline
Signal & \quad $d W^+ c$ \quad & $ d W^+ (\bar s, \bar b) $ & \quad $d W^+ t$  \\ \hline
$\zeta = 0.1$ & 262 & 349 &  64.1  \\
$(S/\sqrt{B})_{\zeta=0.1}$ &  0.34  &  0.45 & 4.0 \\ \hline
$\zeta = 0.2$ & 883 & 1330   & 218    \\
$(S/\sqrt{B})_{\zeta=0.2}$ & 1.1   & 1.7 &  14  \\
 \end{tabular}
 \end{ruledtabular}
\end{table}

Finally, we briefly discuss the new physics in connection to  the flavor physics.  In this study, we do not introduce new couplings to the lepton sector, therefore, the contributions to the lepton flavor-changing processes are similar to the SM predictions. However,  the introduced VLQs lead to FCNCs at the tree level in the quark sector, where the strict constraints  from $\Delta F=2$ processes have been considered  in section III. Besides the rare decays $t\to (u, c)h$ and $t\to (u,c)Z$ that were discussed earlier, it is also interesting to investigate the FCNC effects in the low energy physics. For instance, the coupling $sdZ$ can contribute to the  $K^+\to \pi^+ \nu \bar \nu$ and $K_L \to \pi^0 \nu \bar \nu$ decays, where  the SM predicted BRs are of  ${\cal O}(10^{-11})$, both are sensitive to the new physics effects, and the theoretical uncertainties are well-controlled~\cite{Buras:2015yca}. Furthermore, the NA62 experiment at CERN can achieve the $BR(K^+\to \pi^+ \nu \bar\nu)$ to be a precision of $10\%$~\cite{Rinella:2014wfa,Romano:2014xda}; and the KOTO experiment at J-PARC for $K_L\to \pi^0 \nu \bar\nu$ decays can reach the SM sensitivity.  Thus, it is important to search for new physics in rare $K$ decays. In $B$-meson physics, the tree-level couplings $bqZ$ with $q=d,s$ can contribute to $b\to q \ell^+ \ell^-$ decays. Although the measured $BR(B_s \to \mu^+ \mu^-)$ is consistent with  the SM prediction~\cite{CMS:2014xfa}, a $3.4\sigma$ deviation from the SM prediction in the angular analysis  of $B\to K^* \mu^+ \mu^-$ is observed~\cite{Aaij:2015oid}.  It is worthy to explore the excess in our model. Since the detailed analysis of flavor physics is beyond the scope of this paper,  a complete analysis will be studied elsewhere~\cite{CN}.

\section{Conclusion}

We have studied the phenomenology of which two triplet VLQs with $Y=2/3, -1/3$ and a Higgs singlet are embedded in the SM.  Because the  isospin of VLQs is different from that of the SM quarks,  Higgs- and $Z$-mediated FCNCs are generated at the tree level and the new  CKM matrix becomes a non-unitary matrix. We find that the modifications of the CKM matrix elements coupled to the SM quarks can be smeared out if  two triplet  VLQs are introduced and the scheme $\zeta_{1i} = \zeta_{2i}$ is adopted, where $\zeta_{ij} = v Y_{ij}/m_F$ are the parameters from flavor mixings.

Although the tree-level FCNCs cannot be removed,   it was found that when the constraints from $\Delta F=2$ processes are applied, the upper limits of BRs for $t\to c (h, Z)$ decays  are  $(6.8, 0.48)\times 10^{-5}$, which is  two orders of magnitude smaller than the current experimental bounds.  With the values of constrained parameters, we examined the influence of the model on the SM Higgs production and its diphoton decay; we found that $\sigma(pp\to h)$ and $BR(h\to \gamma\gamma)$ can have $13\%$ and $-2\%$ deviations from the SM results, respectively.  As a result, the signal strength for $pp\to h\to \gamma\gamma$  is thus changed by  $10\%$. 

The main purpose of this work was to explore the single  production of  exotic VLQs $X$ and $Y$ in the $pp$ collisions at $\sqrt{s}=13$ TeV.  We gave a detailed analysis for each possible $q_i q'_j$ scattering, where $q_i$ and $q'_j$ are the possible initial quarks.  It was found that  the contributions of  $s$-channel annihilations are much smaller than those of $t$-channel annihilations. From this study, we comprehend the contribution of each subprocess to the production cross section of a specific VLQ. The interesting production channels are $X_{5/3} d$, $Y_{-4/3} u$, and $Y_{4/3} d$, where the corresponding production cross sections for $m_X=m_Y=1$ TeV are $84.3$, $72.3$, and $157.8$ fb, respectively.  From our analysis, it is clear to see that the single production cross sections of VLQs are much larger than the  pair production cross sections, which are through QCD processes.   The dominant decay modes of the VLQs are $X_{5/3} \to ( c, t) W^+$ and $Y_{-4/3} \to (s, b) W^-$. Each BR can be 1/2 in our chosen scheme. For illustration, we estimate the significances for the channels proposed in Eq.~(\ref{eq:signal}). It is found that the significance for $pp\to d W^+ t$ channel can be over $5\sigma$.  \\

\noindent{\bf Acknowledgments}

 This work was partially supported by the Ministry of Science and Technology of Taiwan
R.O.C.,  under grant MOST-103-2112-M-006-004-MY3 (CHC). 


\begin{thebibliography}{99}
 
 \bibitem{:2012gk} 
  G.~Aad {\it et al.}  [ATLAS Collaboration],
"Observation of a new particle in the search for the Standard Model Higgs
  boson with the ATLAS detector at the LHC", Phys.\ Lett.\ B {\bf 716}, 1 (2012)
  [arXiv:1207.7214 [hep-ex]].

\bibitem{:2012gu} 
  S.~Chatrchyan {\it et al.}  [CMS Collaboration],
"Observation of a new boson at a mass of 125 GeV with the CMS experiment at
  the LHC", Phys.\ Lett.\ B {\bf 716}, 30 (2012)
  [arXiv:1207.7235 [hep-ex]].

\bibitem{Aad:2015owa} 
  G.~Aad {\it et al.} [ATLAS Collaboration],
  JHEP {\bf 1512}, 055 (2015)
  doi:10.1007/JHEP12(2015)055
  [arXiv:1506.00962 [hep-ex]].

   \bibitem{Khachatryan:2014hpa} 
  V.~Khachatryan {\it et al.} [CMS Collaboration],
  JHEP {\bf 1408}, 173 (2014)
  [arXiv:1405.1994 [hep-ex]];
  V.~Khachatryan {\it et al.} [CMS Collaboration],
  JHEP {\bf 1408}, 174 (2014)
  [arXiv:1405.3447 [hep-ex]];
  V.~Khachatryan {\it et al.} [CMS Collaboration],
  Phys.\ Lett.\ B {\bf 740}, 83 (2015)
  [arXiv:1407.3476 [hep-ex]].

\bibitem{Khachatryan:2015kon} 
  V.~Khachatryan {\it et al.} [CMS Collaboration],
  Phys.\ Lett.\ B {\bf 749}, 337 (2015)
  [arXiv:1502.07400 [hep-ex]].



\bibitem{ATLAS-CONF-2015-081} 
  The ATLAS collaboration,
  ATLAS-CONF-2015-081.
  
\bibitem{Aaboud:2016tru} 
  M.~Aaboud {\it et al.} [ATLAS Collaboration],
  arXiv:1606.03833 [hep-ex].
  
\bibitem{CMS:2015dxe} 
  CMS Collaboration [CMS Collaboration],
  CMS-PAS-EXO-15-004.
  
\bibitem{Khachatryan:2016hje} 
  V.~Khachatryan {\it et al.} [CMS Collaboration],
  arXiv:1606.04093 [hep-ex].

\bibitem{Arhrib:2015maa} 
  A.~Arhrib, R.~Benbrik, C.~H.~Chen, M.~Gomez-Bock and S.~Semlali,
  Eur.\ Phys.\ J.\ C {\bf 76}, no. 6, 328 (2016)
  doi:10.1140/epjc/s10052-016-4167-9
  [arXiv:1508.06490 [hep-ph]].
  
  
\bibitem{Eberhardt:2012sb} 
  O.~Eberhardt, G.~Herbert, H.~Lacker, A.~Lenz, A.~Menzel, U.~Nierste and M.~Wiebusch,
  Phys.\ Rev.\ D {\bf 86}, 013011 (2012)
  [arXiv:1204.3872 [hep-ph]].




\bibitem{delAguila:2000rc} 
  F.~del Aguila, M.~Perez-Victoria and J.~Santiago,
  JHEP {\bf 0009}, 011 (2000)
  [hep-ph/0007316].
  
\bibitem{Atre:2008iu} 
  A.~Atre, M.~Carena, T.~Han and J.~Santiago,
  Phys.\ Rev.\ D {\bf 79}, 054018 (2009)
  [arXiv:0806.3966 [hep-ph]].
  
\bibitem{Mrazek:2009yu} 
  J.~Mrazek and A.~Wulzer,
  Phys.\ Rev.\ D {\bf 81}, 075006 (2010)
  [arXiv:0909.3977 [hep-ph]].
  
\bibitem{Cacciapaglia:2010vn} 
  G.~Cacciapaglia, A.~Deandrea, D.~Harada and Y.~Okada,
  JHEP {\bf 1011}, 159 (2010)
  [arXiv:1007.2933 [hep-ph]].
  
\bibitem{Gopalakrishna:2011ef} 
  S.~Gopalakrishna, T.~Mandal, S.~Mitra and R.~Tibrewala,
  Phys.\ Rev.\ D {\bf 84}, 055001 (2011)
  [arXiv:1107.4306 [hep-ph]].
  
\bibitem{Botella:2012ju} 
  F.~J.~Botella, G.~C.~Branco and M.~Nebot,
  JHEP {\bf 1212}, 040 (2012)
  [arXiv:1207.4440 [hep-ph]].

 \bibitem{Okada:2012gy} 
  Y.~Okada and L.~Panizzi,
  Adv.\ High Energy Phys.\  {\bf 2013}, 364936 (2013)
  [arXiv:1207.5607 [hep-ph]].
  
\bibitem{Cai:2012ji} 
  H.~Cai,
  JHEP {\bf 1302}, 104 (2013)
  [arXiv:1210.5200 [hep-ph]].
  
  \bibitem{Cacciapaglia:2012dd} 
  G.~Cacciapaglia, A.~Deandrea, L.~Panizzi, S.~Perries and V.~Sordini,
  JHEP {\bf 1303}, 004 (2013)
  [arXiv:1211.4034 [hep-ph]].
  
\bibitem{Atre:2013ap} 
  A.~Atre, M.~Chala and J.~Santiago,
  JHEP {\bf 1305}, 099 (2013)
  [arXiv:1302.0270 [hep-ph]].
  
\bibitem{Aguilar-Saavedra:2013qpa} 
  J.~A.~Aguilar-Saavedra, R.~Benbrik, S.~Heinemeyer and M.~Perez-Victoria,
  Phys.\ Rev.\ D {\bf 88}, no. 9, 094010 (2013)
  [arXiv:1306.0572 [hep-ph]].
  
\bibitem{Gopalakrishna:2013hua} 
  S.~Gopalakrishna, T.~Mandal, S.~Mitra and G.~Moreau,
  JHEP {\bf 1408}, 079 (2014)
  [arXiv:1306.2656 [hep-ph]].
  
\bibitem{Beauceron:2014ila} 
  S.~Beauceron, G.~Cacciapaglia, A.~Deandrea and J.~D.~Ruiz-Alvarez,
  Phys.\ Rev.\ D {\bf 90}, no. 11, 115008 (2014)
  [arXiv:1401.5979 [hep-ph]].
  
  \bibitem{Alok:2014yua}
  A.~K.~Alok, S.~Banerjee, D.~Kumar and S.~U.~Sankar,
  arXiv:1402.1023 [hep-ph].
  
\bibitem{Karabacak:2014nca} 
  D.~Karabacak, S.~Nandi and S.~K.~Rai,
  Phys.\ Lett.\ B {\bf 737}, 341 (2014)
  [arXiv:1405.0476 [hep-ph]].
     

\bibitem{Cacciapaglia:2015ixa} 
  G.~Cacciapaglia, A.~Deandrea, N.~Gaur, D.~Harada, Y.~Okada and L.~Panizzi,
  JHEP {\bf 1509}, 012 (2015)
  [arXiv:1502.00370 [hep-ph]].
  
  \bibitem{Alok:2015iha}
  A.~K.~Alok, S.~Banerjee, D.~Kumar, S.~U.~Sankar and D.~London,
  Phys.\ Rev.\ D {\bf 92}, no. 1, 013002 (2015)
  [arXiv:1504.00517 [hep-ph]].
  
\bibitem{Vignaroli:2015ama} 
  N.~Vignaroli,
  Phys.\ Rev.\ D {\bf 91}, no. 11, 115009 (2015)
  [arXiv:1504.01768 [hep-ph]].

\bibitem{Chen:2015cfa} 
  C.~H.~Chen and T.~Nomura,
  Phys.\ Rev.\ D {\bf 92}, no. 11, 115021 (2015)
  [arXiv:1509.02039 [hep-ph]].
  
\bibitem{Angelescu:2015kga} 
  A.~Angelescu, A.~Djouadi and G.~Moreau,
  Eur.\ Phys.\ J.\ C {\bf 76}, no. 2, 99 (2016)
  [arXiv:1510.07527 [hep-ph]].
  
\bibitem{Benbrik:2015fyz} 
  R.~Benbrik, C.~H.~Chen and T.~Nomura,
  arXiv:1512.06028 [hep-ph].

  
   
\bibitem{ATLAS:2012qe} 
  G.~Aad {\it et al.} [ATLAS Collaboration],
  Phys.\ Lett.\ B {\bf 718}, 1284 (2013)
  [arXiv:1210.5468 [hep-ex]].
 
\bibitem{Aad:2014efa} 
  G.~Aad {\it et al.} [ATLAS Collaboration],
  JHEP {\bf 1411}, 104 (2014)
  [arXiv:1409.5500 [hep-ex]].
   
\bibitem{Aad:2015kqa} 
  G.~Aad {\it et al.} [ATLAS Collaboration],
  JHEP {\bf 1508}, 105 (2015)
  [arXiv:1505.04306 [hep-ex]].
 
\bibitem{Aad:2015mba} 
  G.~Aad {\it et al.} [ATLAS Collaboration],
  Phys.\ Rev.\ D {\bf 91}, no. 11, 112011 (2015)
  [arXiv:1503.05425 [hep-ex]].
 
\bibitem{Aad:2015gdg} 
  G.~Aad {\it et al.} [ATLAS Collaboration],
  JHEP {\bf 1510}, 150 (2015)
  [arXiv:1504.04605 [hep-ex]].

\bibitem{Aad:2015tba} 
  G.~Aad {\it et al.} [ATLAS Collaboration],
  Phys.\ Rev.\ D {\bf 92}, no. 11, 112007 (2015)
  [arXiv:1509.04261 [hep-ex]].

\bibitem{Aad:2015voa} 
  G.~Aad {\it et al.} [ATLAS Collaboration],
  arXiv:1510.02664 [hep-ex].
     
\bibitem{Aad:2016shx} 
  G.~Aad {\it et al.} [ATLAS Collaboration],
  arXiv:1602.06034 [hep-ex].

\bibitem{Chatrchyan:2013uxa} 
  S.~Chatrchyan {\it et al.} [CMS Collaboration],
  Phys.\ Lett.\ B {\bf 729}, 149 (2014)
  [arXiv:1311.7667 [hep-ex]].

\bibitem{Chatrchyan:2013wfa} 
  S.~Chatrchyan {\it et al.} [CMS Collaboration],
  Phys.\ Rev.\ Lett.\  {\bf 112}, no. 17, 171801 (2014)
  [arXiv:1312.2391 [hep-ex]].
  
\bibitem{Khachatryan:2015gza} 
  V.~Khachatryan {\it et al.} [CMS Collaboration],
  arXiv:1507.07129 [hep-ex].

\bibitem{Khachatryan:2015oba} 
  V.~Khachatryan {\it et al.} [CMS Collaboration],
  Phys.\ Rev.\ D {\bf 93}, no. 1, 012003 (2016)
  [arXiv:1509.04177 [hep-ex]].

\bibitem{Khachatryan:2015axa} 
  V.~Khachatryan {\it et al.} [CMS Collaboration],
  JHEP {\bf 1506}, 080 (2015)
  [arXiv:1503.01952 [hep-ex]].
    
\bibitem{PDG2014}
K.A. Olive et al. (Particle Data Group), Chin. Phys. C, 38, 090001 (2014).      
 
\bibitem{Buras:2001ra} 
  A.~J.~Buras, S.~Jager and J.~Urban,
  Nucl.\ Phys.\ B {\bf 605}, 600 (2001)
  [hep-ph/0102316].
 
 
\bibitem{Buras:2012fs} 
  A.~J.~Buras and J.~Girrbach,
  JHEP {\bf 1203}, 052 (2012)
  [arXiv:1201.1302 [hep-ph]].
  
\bibitem{Aad:2015gba} 
  G.~Aad {\it et al.} [ATLAS Collaboration],
  arXiv:1507.04548 [hep-ex].
  
  \bibitem{CMS} S. Chatrchyan et al. . [CMS Collaboration], CMS-PAS-HIG-14-009

\bibitem{Gunion:1989we} 
  J.~F.~Gunion, H.~E.~Haber, G.~L.~Kane and S.~Dawson,
  Front.\ Phys.\  {\bf 80}, 1 (2000).

\bibitem{Aad:2015pja} 
  G.~Aad {\it et al.} [ATLAS Collaboration],
  arXiv:1509.06047 [hep-ex].
  
\bibitem{CMStqh}
CMS Collaboration, CMS-PAS-TOP-14-019 (2015).

\bibitem{Aad:2015uza} 
  G.~Aad {\it et al.} [ATLAS Collaboration],
  arXiv:1508.05796 [hep-ex].

\bibitem{Chatrchyan:2013nwa} 
  S.~Chatrchyan {\it et al.} [CMS Collaboration],
  Phys.\ Rev.\ Lett.\  {\bf 112}, no. 17, 171802 (2014)
  [arXiv:1312.4194 [hep-ex]].
  
  \bibitem{Belyaev:2012qa} 
  A.~Belyaev, N.~D.~Christensen and A.~Pukhov,
  Comput.\ Phys.\ Commun.\  {\bf 184}, 1729 (2013)
  [arXiv:1207.6082 [hep-ph]].
  
\bibitem{Nadolsky:2008zw} 
  P.~M.~Nadolsky, H.~L.~Lai, Q.~H.~Cao, J.~Huston, J.~Pumplin, D.~Stump, W.~K.~Tung and C.-P.~Yuan,
  Phys.\ Rev.\ D {\bf 78}, 013004 (2008)
  [arXiv:0802.0007 [hep-ph]].
  
  
\bibitem{Alwall:2014hca} 
  J.~Alwall {\it et al.},
  JHEP {\bf 1407}, 079 (2014)
  [arXiv:1405.0301 [hep-ph]].

  
  \bibitem{Alloul:2013bka} 
  A.~Alloul, N.~D.~Christensen, C.~Degrande, C.~Duhr and B.~Fuks,
  Comput.\ Phys.\ Commun.\  {\bf 185}, 2250 (2014)
  [arXiv:1310.1921 [hep-ph]].
  
\bibitem{Deans:2013mha} 
  C.~S.~Deans [NNPDF Collaboration],
  arXiv:1304.2781 [hep-ph].
  
  \bibitem{Ref:Pythia}
  T.~Sjostrand, S.~Mrenna, P.~Z.~Skands,
  JHEP {\bf 0605 },  026 (2006).



\bibitem{Ref:PGS}
\url{http://www.physics.ucdavis.edu/conway/research/software/pgs/pgs4-general.htm}.

\bibitem{Ball:2007zza} 
  G.~L.~Bayatian {\it et al.}  [CMS Collaboration],
  J.\ Phys.\ G {\bf 34}, 995 (2007).


\bibitem{Buras:2015yca} 
  A.~J.~Buras, D.~Buttazzo and R.~Knegjens,
  JHEP {\bf 1511}, 166 (2015)
  [arXiv:1507.08672 [hep-ph]].
  
  \bibitem{Rinella:2014wfa} 
  F.~Newson {\it et al.},
  arXiv:1411.0109 [hep-ex].
  
  \bibitem{Romano:2014xda} 
  A.~Romano,
  arXiv:1411.6546 [hep-ex].
  
  \bibitem{CMS:2014xfa} 
  V.~Khachatryan {\it et al.} [CMS and LHCb Collaborations],
  Nature {\bf 522}, 68 (2015)
  [arXiv:1411.4413 [hep-ex]].
  
  \bibitem{Aaij:2015oid} 
  R.~Aaij {\it et al.} [LHCb Collaboration],
  JHEP {\bf 1602}, 104 (2016)
  [arXiv:1512.04442 [hep-ex]].
  
  \bibitem{CN}C.H. Chen and T. Nomura, study of rare $K$, $D$, and $B$ meson decays in this model is in progress. 

\end{thebibliography}
\end{document}